%% file: DistCov.tex
\documentclass[sigconf,screen]{acmart}

\usepackage{enumitem}
\usepackage{algorithm}
\usepackage{algorithmic}
\usepackage{tikz}
\usepackage{amsmath,bm}
\usepackage{multirow}
\usepackage{booktabs} % For formal tables
\usepackage{threeparttable}
\usepackage[caption=false]{subfig}

\usepackage{array}
\usepackage{hhline}

\DeclareMathOperator{\Expect}{\mathbb{E}}
\DeclareMathOperator{\Var}{\mathrm{Var}}

\newtheorem*{problem}{Problem}

\newcommand{\njuaffl}[0]{%
\affiliation{%
  \institution{State Key Lab of Novel Software Technology, Nanjing University}%
  \city{Nanjing}%
  \country{China}%
}%
}

\setcopyright{rightsretained}
\acmPrice{}
\acmDOI{10.1145/3338906.3338930}
\acmYear{2019}
\copyrightyear{2019}
\acmISBN{978-1-4503-5572-8/19/08}
\acmConference[ESEC/FSE '19]{Proceedings of the 27th ACM Joint European Software Engineering Conference and Symposium on the Foundations of Software Engineering}{August 26--30, 2019}{Tallinn, Estonia}
\acmBooktitle{Proceedings of the 27th ACM Joint European Software Engineering Conference and Symposium on the Foundations of Software Engineering (ESEC/FSE '19), August 26--30, 2019, Tallinn, Estonia}

\begin{document}
\title{Boosting Operational DNN Testing Efficiency through Conditioning}
%\titlenote{Produces the permission block, and
%  copyright information}
%\subtitle{Extended Abstract}
%\subtitlenote{The full version of the author's guide is available as
%  \texttt{acmart.pdf} document}

\author{Zenan Li}%
\njuaffl
\email{lizenan@smail.nju.edu.cn}
\author{Xiaoxing Ma}
\authornote{Corresponding author.}
\orcid{0000-0001-7970-1384}
\njuaffl
\email{xxm@nju.edu.cn}
\author{Chang Xu}
\njuaffl
\email{changxu@nju.edu.cn}
\author{Chun Cao}
\njuaffl
\email{caochun@nju.edu.cn}
\author{Jingwei Xu}
\njuaffl
\email{jingweix@nju.edu.cn}
\author{Jian L\"{u}}
\njuaffl
\email{lj@nju.edu.cn}

%% The default list of authors is too long for headers.

\begin{abstract}
With the increasing adoption of Deep Neural Network (DNN) models as integral parts of software systems, 
efficient operational testing of DNNs is much in demand to ensure these models' actual performance in field conditions. 
A challenge is that the testing often needs to produce precise results with a very limited budget 
%of expensive labeled data.  
for labeling data collected in field.

Viewing software testing as a practice of reliability estimation through statistical sampling, 
we re-interpret the idea behind conventional structural coverages as conditioning for variance reduction.  
With this insight we propose an efficient DNN testing method based on the conditioning on
the representation learned by the DNN model under testing. 
The representation is defined by the probability distribution of the output of neurons 
in the last hidden layer of the model.
To sample from this high dimensional distribution in which the operational data are sparsely distributed, 
we design an algorithm leveraging cross entropy minimization.

Experiments with various DNN models and datasets were conducted to evaluate the general efficiency of the approach. 
The results show that, compared with simple random sampling, 
this approach requires only about a half of labeled inputs to achieve the same level of precision. 

\end{abstract}

\begin{CCSXML}
<ccs2012>
<concept>
<concept_id>10011007.10011074.10011099.10011102.10011103</concept_id>
<concept_desc>Software and its engineering~Software testing and debugging</concept_desc>
<concept_significance>500</concept_significance>
</concept>
<concept>
<concept_id>10010147.10010257.10010293.10010294</concept_id>
<concept_desc>Computing methodologies~Neural networks</concept_desc>
<concept_significance>500</concept_significance>
</concept>
</ccs2012>
\end{CCSXML}

\ccsdesc[500]{Software and its engineering~Software testing and debugging}
\ccsdesc[500]{Computing methodologies~Neural networks}

\keywords{Software testing, Neural networks, Coverage criteria}

\maketitle

\input{introduction}
\input{coverage}
\input{method}

\input{evaluation}

\input{relatedwork}

\input{conclusions}

\begin{acks}
   The authors would like to thank the anonymous reviewers for their suggestions. 
   This work is supported by the National Key R\&D Program of China (2017YFB1001801), the NSFC (61690204, 61802170), and the Collaborative Innovation Center of Novel Software Technology and Industrialization.
\end{acks}

\bibliographystyle{ACM-Reference-Format}
\bibliography{DistCov}

\end{document}

%% file: introduction.tex
% !TEX root = DistCov.tex

\section{Introduction}

Deep Learning has gained great success in tasks that are intuitive to human but hard to describe formally, 
such as image classification or speech recognition~\cite{lecun2015Nature-DeepLearning,Goodfellow-2016-DL}.  
As a result, Deep Neural Networks (DNNs) are increasingly adopted as integral parts of widely used software systems, 
including those in safety-critical application scenarios such as medical diagnosis~\cite{Obermeyer-2016-NEJM} and self-driven cars~\cite{Pei_2017_SOSP}. 
Effective and efficient testing methods for DNNs are thus needed to ensure their service quality in operation environments.

Recent efforts on DNN testing~\cite{DeepGaugeASE18,Pei_2017_SOSP,DBLP:journals/corr/abs-1803-04792,ma2018combinatorial,sun2018concolic,odena2018tensorfuzz,zhang2018deeproad} 
have aimed at generating artificial adversarial examples, 
which resembles the \emph{debug testing}~\cite{Frankl_1998_TES} of human written programs 
that aims at finding error-inducing inputs. 
However, the fundamental difference between DNN models and human written programs
challenges the basic concepts and wisdoms for debug testing. 
For example, the inductive nature of statistical machine learning and the No-Free-Lunch theorem imply 
that an oracle for a DNN model independent of its operation context is senseless. 
The fact that DNN performance is measured statistically also diminishes the importance of individual error-inducing inputs.

%Contrasting to recent efforts on DNN testing aimed at generating artificial adversarial 
%examples~\cite{DeepGaugeASE18,Pei_2017_SOSP,DBLP:journals/corr/abs-1803-04792,ma2018combinatorial,sun2018concolic,odena2018tensorfuzz,zhang2018deeproad}, 
%%which borrows wisdoms from conventional software testing such as 
%this paper focuses on \emph{operational DNN testing}, 
%i.e., testing a previously trained DNN model with the data collected from a specific operation context, 
%in order to determine the model's \emph{actual} performance in this context. 
%%
%The rationale comes from the fundamental difference between DNN models and  
%%However, DNNs are fundamentally different from 
%human written programs, 
%which challenges the basic concepts and wisdoms for conventional software testing
%that aims at finding error-inducing inputs (called \emph{debug testing}~\cite{Frankl_1998_TES}). 
%For example, the inductive nature of statistical machine learning and the No-Free-Lunch theorem imply 
%that an oracle for a DNN model independent of its operation context is senseless. 
%The fact that DNN performance is measured statistically also diminishes the importance of individual error-inducing inputs. 

Contrastingly, this paper focuses on the \emph{operational testing} of DNN, i.e., 
testing a previously trained DNN model with the data collected from a specific operation context, 
in order to determine the model's \emph{actual} performance in this context. 
Although operational testing for conventional software has been extensively studied~\cite{Musa-1993-IEEESoftware,Frankl_1998_TES,Lyu-2007-SRE}, 
%and some recent proposals make wise combinations of operational and debug testing~\cite{Cotr_2016_TSE,Bertolino_2017_ICSE},
the challenge of operational DNN testing is not well understood in the software engineering community. 
A central problem here is that it can be prohibitively expensive to label all the operational data collected in field. 
For example, a surgical biopsy may be needed to decide whether a radiology or pathology image is really malignant or benign. 
In this case the labeling effort for each single example is worth saving. 
Thus it is crucial to test DNN \emph{efficiently}, 
i.e., to precisely estimate a DNN's actual performance in an operation context, 
but with a limited budget for labeling data collected from this context. 

We propose to reduce the number of labeled examples required in operational DNN testing 
through carefully designed sampling.
The conventional wisdom behind structural coverages for testing human written programs 
is re-interpreted in statistical terms as conditioning for variance reduction, 
and applied to the sampling and estimation of DNN's operational accuracy.  

The key insight is that, the representation learned by a DNN and encoded in the neurons in the last hidden layer 
can be leveraged to guide the sampling from the unlabeled operational data. 
It turns out that conditioning on this representation is effective, 
and works well even when the model is not well-fitted to the operation data, 
which is a property not enjoyed by naive choices such as stratifying by classification confidence.   

To realize the idea, one must select a small fraction from the operational data,
but with sufficient representativeness in terms of their distribution in the 
space defined by the outputs of neurons in the last hidden layer. 
This is difficult because the space is high-dimensional, 
and in which the operational data themselves are sparsely distributed. 
We solve this problem with a distribution approximation technique based on cross-entropy minimization.

The contributions of this paper are: 
\begin{itemize}
	\item A formulation of the problem of operational DNN testing as the estimation of performance with
			a small sample, and a proposal for efficient testing with variance reduction 
			through conditioning, as a generalization of structural coverages. 
	\item An efficient approach to operational DNN testing 
			that leverages the high dimensional representation learned by the DNN under testing, 
			and a sampling algorithm realizing the approach based on cross entropy minimization. 
	\item A systematic empirical evaluation. 
			Experiments with LeNet, VGG, and ResNet show that, compared with simple random sampling, 
			this approach requires only about a half of labeled inputs to achieve the same level of precision. 
\end{itemize}

The rest of this paper is organized as follows. 
In Section~\ref{sec:TnCRvstd} we discuss the problem of operational DNN testing and how to improve its efficiency. 
% in the sense of reducing the requirement on labeled operational data.  
Section~\ref{sec:TstDNN} is devoted to the conditioning approach to efficient DNN testing, and 
%analyzes problems in previous coverage criteria.
Section~\ref{sec:Evaluation} to the empirical evaluation of the approach.
%explains an effective coverage criteria which allow us to evaluate the testing set.
We then %discuss some tricky issues and possible variances of our approach in Section~\ref{sec:Discussions}, 
%and 
review related work in Section~\ref{sec:RelatedWork}, before concluding the paper with Section~\ref{sec:Conclusions}.  
%shows our experiments and Section 6 concludes.

%% file: coverage.tex
% !TEX root = DistCov.tex

\section{Operational Testing of DNNs}
\label{sec:TnCRvstd}

In this section we briefly introduce DNN, examine the problem of testing DNNs as software artifacts in operation context, 
and then discuss the insights for and the challenges to efficient DNN testing. 

%\subsection{DNNs as software components}
\subsection{Deep Neural Network}
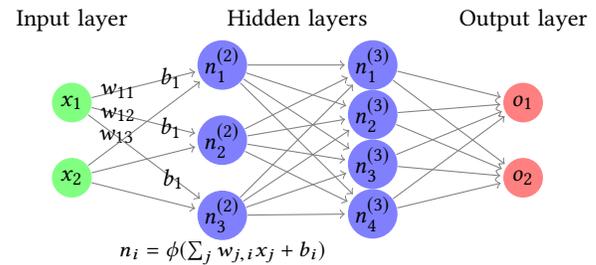
\begin{figure}[h]
\begin{center}
\def\layersep{2cm}
\begin{tikzpicture}[shorten >=1pt,->,draw=black!50, node distance=\layersep]
    \tikzstyle{every pin edge}=[<-,shorten <=1pt]
    \tikzstyle{neuron}=[circle,fill=black!25,minimum size=15pt,inner sep=0pt]
    \tikzstyle{input neuron}=[neuron, fill=green!50];
    \tikzstyle{output neuron}=[neuron, fill=red!50];
    \tikzstyle{hidden neuron}=[neuron, fill=blue!50];
    \tikzstyle{annot} = [text width=6em, text centered]

    % Draw the input layer nodes
    \foreach \name / \y in {1,...,2}
    % This is the same as writing \foreach \name / \y in {1/1,2/2,3/3,4/4}
        \node[input neuron] (I-\name) at (-1.5* \layersep,-\y) {$x_\y$};

    % Draw the first hidden layer nodes
    \foreach \name / \y in {1,...,3}
        \path[yshift=0.5cm]
            node[hidden neuron] (H1-\name) at (-0.5* \layersep,-\y cm) {$n_{\y}^{(2)}$};
            
    % Draw the second hidden layer nodes
    \foreach \name / \y in {1,...,4}
        \path[yshift=0.16cm]
            node[hidden neuron] (H2-\name) at (0.5*\layersep,- 0.66* \y cm) {$n_{\y}^{(3)}$};

    % Draw the output layer node
    \foreach \name / \y in {1,...,2}
    \node[output neuron] (O-\name) at (1.5*\layersep,-\y cm) {$o_\y$};

    % Connect every node in the input layer with every node in the
    % hidden layer.
    \foreach \source in {1,...,1}
        \foreach \dest  / \y in {1,...,3}
            \path (I-\source) edge node[near start] {$w_{1{\y}}$} node[near end]{$b_1$} (H1-\dest);
            
    \foreach \source in {2,...,2}
        \foreach \dest  in {1,...,3}
            \path (I-\source) edge  (H1-\dest);

    % Connect every node in the hidden layer 
    \foreach \source in {1,...,3}
            \foreach \dest in {1,...,4}
             \path (H1-\source) edge (H2-\dest);
             
    % Connect every node in the hidden layer with the output layer
    \foreach \source in {1,...,4}
            \foreach \dest in {1,...,2}
             \path (H2-\source) edge (O-\dest);

    % Annotate the layers
   \node[annot,above of=H2-1, node distance=0.6cm] (hl) {};
   \node[annot,left of =hl, node distance=1cm] {Hidden layers};
    \node[annot,above of=I-1, node distance=1.1cm] {Input layer};
    \node[annot,above of=O-1, node distance=1.1cm] {Output layer};
    
    %\draw[dashed, color=black, line width=0.5pt] (-0.3 * \layersep,-2.8cm) node[below]{$n_i=\phi(\sum_{j}w_{j,i}x_j + b_i)$} ->(-0.5 * \layersep,-2.8cm);
    \node[below of=H1-3, node distance=0.5cm]{\small{$n_i=\phi(\sum_{j}w_{j,i}x_j + b_i)$}};

\end{tikzpicture}
\end{center}
\caption{A simple neural network}
\label{fig:DNN}
\end{figure}

A deep neural network (DNN) is an artificial neural network (ANN) with multiple intermediate (hidden) layers.
%between the input and output layers. 
%The DNN finds the correct mathematical manipulation to turn inputs into their corresponding outputs, 
%whether it is a linear relationship or a non-linear relationship. 
It encodes a mathematical mapping from inputs to outputs with a cascading composition of simple functions implemented by the neurons. % which approximates the distribution hidden in the training data, 
%Concretely, the inputs of a next layer is the linear combinations of outputs of its previous layer, and the existence of activation functions $\phi$ makes the model nonlinear. 
Figure~\ref{fig:DNN} is a simple example of  neural network. The existence of activation functions $\phi$ makes the model nonlinear. 

%Similar to other machine learning models, a DNN model is learned from a training dataset, 
To approximate the intricate mapping hidden in the training examples, 
a DNN model has its parameters (weights $w_{i,j}$ and biases $b_i$) 
gradually adjusted to minimize the averaged prediction error over all the examples.
%which is a set of examples used to fit the parameters ().
What a DNN actually learned is a posterior probability distribution, denoted as $p(y \mid \bm{x})$.
For example, for a \textit{k}-classification problem, DNN will give \textit{k} posterior probability functions $p(y = i \mid \bm{x}), i=1,2,\dots,k$ for the given input $\bm{x}$.
The predicted label for this input is the class corresponding to the maximum posteriori probability, i.e. $f(\bm{x}) = \arg\max_i p(y = i \mid \bm{x})$.

%Normally, DNN testing is limited to the accuracy of the test set.
%For detail, test dataset is used to provide an unbiased evaluation of a trained model.
%In this sense, the accuracy of test dataset describes the difference between the posterior distribution of model learning and the actual posterior distribution.
%At the same time, the reliability of accuracy depends on the distribution of test sets, i.e. the distribution of test sets is expected to be similar to the real task distribution, so that accuracy can effectively characterize the overall performance of the model in practical applications.

%In this paper, we assume that if the test set is large enough, the distribution of examples in the test set is consistent with that in the real task.
%However, when the size of test set is small, the inadequacy of test samples will bring unstablility of the accuracy.

\subsection{The DNN Testing Problem}
\label{subsec:DNNTesting}

When a previously trained DNN model is adopted as an integral part of a software system 
deployed in a specific environment, it may drastically underperform its expected accuracy.
There can be different causes, such as under-fitting or over-fitting of the model to the training data set, 
or the data distribution discrepancy between the training set and the data emerged in the operation context. 
The latter is especially nasty and often encountered in practice. 
Therefore, as any software artifact, a DNN model must be sufficiently tested before being put into production.

DNN testing is different from traditional software testing aiming at identifying error-inducing inputs.
%\footnote{%
%	To be fair, there are different purposes of testing machine learning programs~\cite{Houssem-2018-TestMLProg}, 
%	including those searching for adversarial inputs~\cite{DeepGaugeASE18,Pei_2017_SOSP,DBLP:journals/corr/abs-1803-04792,ma2018combinatorial,sun2018concolic,odena2018tensorfuzz,zhang2018deeproad}.
%	There are also reliability estimation through operational testing for conventional software
%	systems~\cite{Frankl_1998_TES,Lv_2014_TSE,Cotr_2016_TSE}. 
%	We will discuss them later in Section~\ref{sec:RelatedWork}.}. 
DNN implements a kind of inductive reasoning, 
which is fundamentally different from human written programs based on logic deductions. 
As a consequence, for a trained DNN there does not exist a certain and universal oracle for testing. 
Elaborately, the testing of DNNs has to be 
\begin{description}
	\item[Statistical] 
	Contrasting to human written programs with certain intended behaviors, 
	as a statistical machine learning model, a DNN offers only some probabilistic guarantee, 
	i.e., to make \emph{probably} correct prediction on \emph{most} inputs it concerns~\cite{Goodfellow-2016-DL}.
	In fact, mispredictions on a small portion of inputs are \emph{expected}, 
	and in some sense \emph{intentional}, in order to avoid overfitting and maximizing generality. 
	\item[Holistic] 
	Up to now there is no viable rationale interpretation of 
	DNNs' internal behaviors on individual inputs at the level of neurons~\cite{Zou-2019-AAAI-Fragile,Lipton-2016-arXiv-Mythos}.
	This means that DNNs are essentially blackboxes although their computation steps are visible. 
	Also, a detected fault with a specific input is hardly helpful for ``debugging'' the DNN.    
	\item[Operational] 
	Moreover, testing of DNNs without considering their operation context is meaningless. 
	This is implied by the No Free-Lunch Theorem~\cite{Wolpert_1996_NFL}, 
	which says that, considering all possible contexts, no machine learning algorithm is 
	universally any better than any other~\cite{Goodfellow-2016-DL}. 
\end{description}

So generally the task of testing a DNN as a software component is, 
giving a previously trained DNN model and a specific operation context, 
to decide how well the model will perform in this context, 
which is expressed statistically with the estimated accuracy of prediction%
\footnote{In this paper we consider only accuracy that is the proportion of examples for which the model predicts correctly. 
However, the proposed method is generally applicable to other performance measures.}. 
This task should be easy if we had enough \emph{labeled} data 
that well represent the operation context and suffice accurate estimation.  
However, in practice, although unlabeled data can be collected from the operation environment, 
labeling them with high-quality is often expensive. % due to the human expertise required.   

For example, considering an application scenario of AI-aided clinical medicine~\cite{Obermeyer-2016-NEJM} 
where a hospital is going to adopt a DNN model to predict MRI images to be malignant or benign. 
Suppose that the model is previously trained by a foreign provider with its proprietary dataset, 
and thus the hospital needs to gauge it against native patients and local equipment settings. 
The hospital may collect a lot of images by scanning patients and volunteers, 
but labeling them is much more expensive because not only advanced human expertise,  
but also some complicated laboratory testings and even intrusive biopsies are required.

Therefore, a central problem of DNN testing is how to accurately estimate 
DNNs' performance in their operational context with small-size samples%
\footnote{There is a common mistake of regarding a sample of more than 30 elements as large enough~\cite{Beleites-2013-ACA}.
%In addition, the widely used 95\% standard confidence interval is also often insufficient~\cite{Beleites-2013-ACA}.
As we will see in Section~\ref{sec:Evaluation}, we often need much more. % hundreds of inputs to get an accurate estimation. 
}
%  If 90 out of 100 samples of a class are recognized correctly (e.g. sensitivity of the leukocytes with 25 training samples), 
% the 95% confidence interval for the sensitivity ranges from 0.83 to 0.94 – 
% which in the context of our classification task reads as being between “quite bad” and “really good”.,
 of labeled data. 
Or in other words, 
\emph{given a budget of cost in labeling examples, how to make the estimation of a DNN's performance as accurate as possible}. 

\begin{figure}[htbp]
\begin{center}
\includegraphics[width=0.9\columnwidth]{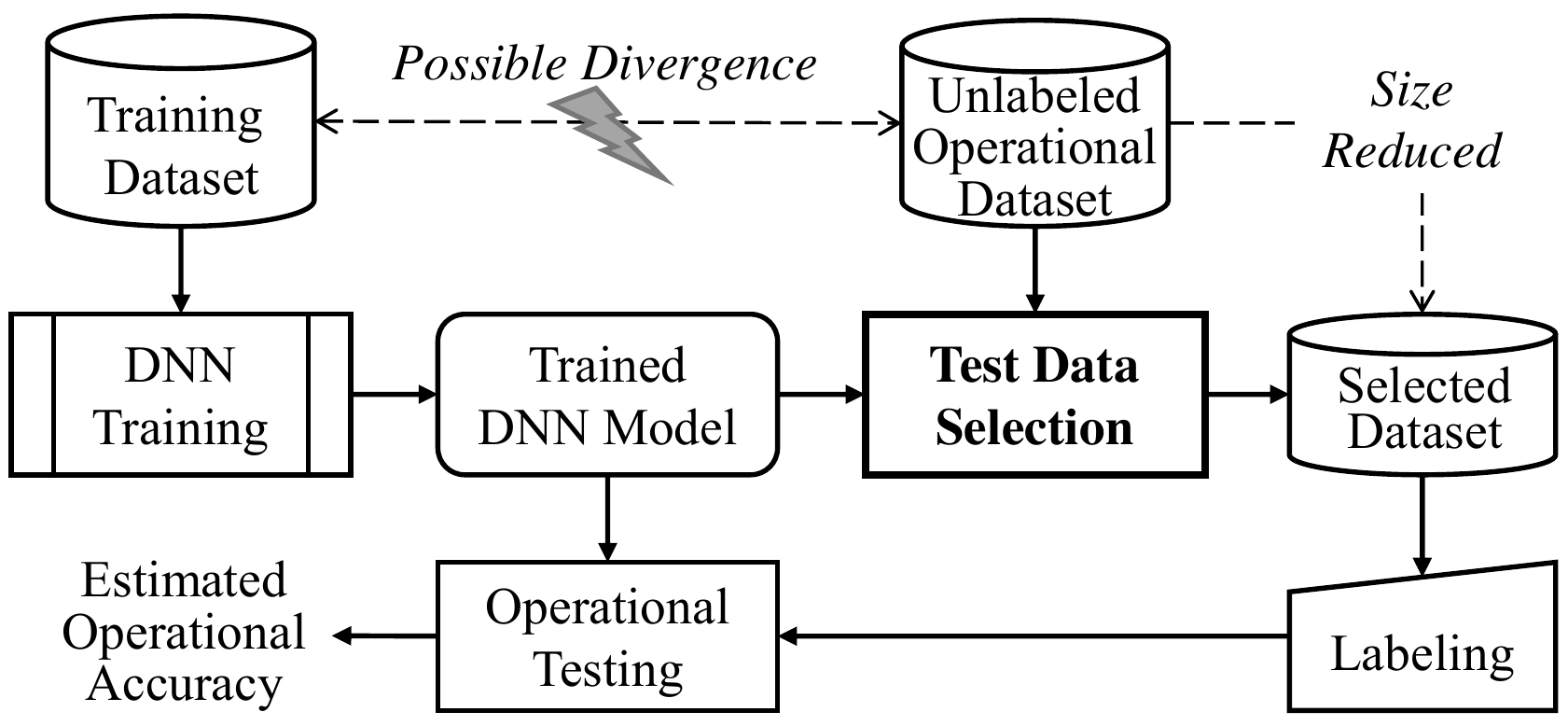}
\caption{Efficient operational DNN testing}
\label{fig:process}
\end{center}
\end{figure}

Figure~\ref{fig:process} illustrates the process of efficient operational DNN testing. 
The goal is that, with some sophisticated test data selection, 
one only needs to label a small portion of operational data to achieve enough precision for the estimation of operational accuracy. 
 
% Obviously, this is problem of statistical sampling. 

\subsection{Improving Testing Efficiency through Conditioning}
\label{subsec:conditioning}

Clearly, the above description of DNN testing resembles statistical sampling and estimation, 
whose efficiency can be improved with variance reduction techniques~\cite{Owen-2013-mcbook}. 
In the following we briefly introduce the simple random sampling as the baseline estimation method, 
and then discuss how conditioning can help with some inspirations from 
the coverage-oriented testing of conventional programs.

%\subsubsection{Distribution coverage}
%Distribution coverage: First, distribution approximation
%can be viewed as a form of implicit stratification that is close to stratified sampling with proportional allocation. 
%In this section, we first give an outline of sampling and
%estimation theories.
%Afterwards, we provide a simple introduction about Deep Neural Network and the purpose of the test set in deep learning algorithm.
%Finally, we briefly discusse some current coverage-guided testing methods for deep learning systems.

% \subsection{Sampling and Estimation Theories}
% Suppose we want to study some property of population, but the total population size is too large to study each individuals. 
% Therefore, sampling is a feasible and effective method.

Sampling is the process of selecting a group of individuals from a population in order to study them and estimate the property of population.
% The accuracy in machine learning is a typical example. 
% We want to estimate the classification performance of the program for the current learning task as a whole. 
% So we choose a part of the samples (i.e. test examples) to estimate the overall performance through the performance of this part of the samples.
% 被估量, 估计量, 估计值.
Specifically, suppose that there is a fixed parameter $\theta = \Expect[H(\bm{X})]$ that needs to be estimated, where $\bm{X}:\Omega \rightarrow \mathcal{D} \subset {\mathbb{R}}^d$ is a random variable corresponding to the observed data, 
and $H:\mathcal{D} \rightarrow \mathbb{R}$ is the model of interest. 
For example, when estimating the accuracy of a DNN model, $H$ is defined as 
$H(\bm{x})=1$ if the DNN correctly predicts $\bm{x}$'s label, and $0$ otherwise.
Note that $H(\bm{X})$ is also a random variable.
% Then an "estimator" is a function that maps the sample space to a set of sample estimates. 
An estimator of $\theta$  is denoted by the symbol $\hat{\theta}$.  
%The estimate for a particular observed data value $x$ (i.e. for $X=x$) is denoted by $H(\bm{X})$, which is a fixed value. 

%Furthermore, different sampling strategies have different estimation methods for estimator.
%Here we introduce two simple sampling strategies.

%\begin{itemize}[leftmargin=*]
%\item[1.] 

The basic sampling method is \emph{Simple Random Sampling} (SRS). 
SRS draws i.i.d.\ replications $x_1,\dots,x_n$ directly from the population.
The i.i.d. condition requires that each individual is chosen randomly and entirely by chance, 
such that each individual has the same probability of being chosen, 
and each set of $n$ individuals has the same probability of being chosen for the sample as any other set of $n$ individuals.
And the estimator $\hat{\theta}$ is computed as the average of each estimate:
\begin{equation}
\hat{\theta} = \frac{H(x_1) + \dots + H(x_n)}{n}\,.
\end{equation}
This is an unbiased estimator, i.e. $\Expect[\hat{\theta}] = \theta$.
The efficiency of SRS is expressed statistically by its variance $\Var[\hat{\theta}] = \frac{1}{n}{\Var[H(\bm{X})]}$.

Despite of its simplicity, SRS is quite effective in practice if we can draw a sufficiently large i.i.d.\ sample. 
Without further information about $H(\bm{X})$ we can hardly improve over SRS. 
An often used strategy is \emph{conditioning}, i.e., to find a random variable or vector $\bm{Z}$, 
on which ideally $H(\bm{X})$ strongly depends, and leverage the law of total variance:
\begin{equation}
\label{equ:ltv}
\Var[H(\bm{X})] = \Expect[\Var[H(\bm{X}|\bm{Z})]] + \Var[\Expect[H(\bm{X}|\bm{Z})]]\,. 
\end{equation}

Intuitively, the law says if we ``interpret'' $H(\bm{X})$ with $\bm{Z}$, 
the variance of $H(\bm{X})$ can be decomposed to those not explained by $\bm{Z}$ (the first term on the right hand), 
and those due to $\bm{Z}$ (the second term). 
Note that $\Expect[H(\bm{X}|\bm{Z})]$ itself is a function on $\bm{Z}$, and 
\begin{equation}
\label{equ:expcond}
\Expect[H(\bm{X})] = \Expect[\Expect[H(\bm{X}|\bm{Z}])]\,.
\end{equation}
So we can sample from $\bm{Z}$'s distribution and estimate $\Expect[H(\bm{X}|\bm{Z})]$ instead of $H(\bm{X})$, 
taking the advantage that the former has a smaller variance than the latter. 

If we can make a complete sample of $\bm{Z}$, i.e., covering all possible values $\bm{z_i}$ for $\bm{Z}$, 
the variance of our estimation will be only those introduced in estimating $\Expect[H(\bm{X}|\bm{z_i})]$ for each $\bm{z_i}$. 
This is exactly what \emph{Stratified Sampling} does. Furthermore, if the value of $H(\bm{X})$ is fully determined by $\bm{Z}$,
we will have zero variance. 

These two ideal conditions are hard to satisfy, especially in complex scenarios such as operational software testing. 
However, the  insights are clear: 
\emph{to improve the efficiency of testing as estimation, we need}
\begin{enumerate}
\item \emph{to identify an observable factor $\bm{Z}$ that affects $H(\bm{X})$ the performance (accuracy) as much as possible, 
so that the variance of $H(\bm{X})$ conditioned under each $\bm{z_i}$ is minimized, and }
\item \emph{to draw as representative as possible samples for $\bm{Z}$, so that the uncertainty due to  $\bm{Z}$ can be well handled.}
\end{enumerate}
These two aspects can be conflicting in practice. Intuitively, the more ``precise'' $\bm{Z}$ the interpretation for $H(\bm{X})$ is,
the finer grained it has to be, and the harder it can be sufficiently represented by a small-size sample. 
It is crucial to strike a good balance between them with a deliberately chosen $\bm{Z}$.

\smallskip
It is inspiring to use this viewpoint to examine the structural coverage-directed testing of conventional programs, 
despite of the difference that structural coverages are mainly used to identify error-inducing inputs. 
The efficacy of a structural coverage comes from  
\begin{enumerate}
\item the \emph{homogeneity} of inputs covering the same part of a program -- either all or none of them induces an error, so that testing efficiency can be improved by avoiding duplications, 
and 
\item the \emph{diversity}  of inputs indicated by coverage metrics, so that testing completeness can be improved by enforcing a high coverage to touch corner cases in which rare errors may hide. 
\end{enumerate}
Evidently, these two heuristics resemble the two insights of conditioning. 

\smallskip
In this sense, the conditioning techniques for testing efficiency improvement can be regarded as 
a generalization of structural coverages in conventional white-box software testing. 
However, it turns out to be challenging to apply this idea to the testing of DNNs, because of
\begin{enumerate}
\item the \emph{blackbox nature of DNNs}. There is no obvious structural features 
like program branches or execution paths of human written programs 
that intuitively provide the needed homogeneity, and moreover,
\item the \emph{curse of dimensionality}. 
For DNNs, a powerful interpreting factor $\bm{Z}$ for model accuracy is often a high dimensional vector, 
which makes it very difficult to represent with a small-size sample but without huge uncertainty.  
\end{enumerate} 
We will discuss how to meet these challenges in the next section. 

%% file: method.tex
% !TEX root = DistCov.tex

\section{Efficient DNN testing methods} %guided by distribution coverage}
\label{sec:TstDNN}

First, let us state our problem of efficient operational DNN testing more specifically:
\begin{problem} 
Given $\mathfrak{M}$ a trained DNN model, and $S$ a set of $N$ unlabeled examples collected from an operation context, 
instead of labeling all these $N$ examples, select and label a subset $T \subseteq S$ with a given size budget $n=|T| \ll N$, 
and use $T$ to estimate the accuracy of \hspace{0em} $\mathfrak{M}$ on $S$, with an as small estimation error as possible.
\end{problem}

Leveraging the information provided by $\mathfrak{M}$ and $S$, we try to achieve efficient estimation through conditioning. 
We first discuss \emph{Confidence-based Stratified Sampling} (CSS), 
which is simple but unfortunately fragile and limited to classifiers. 
Then we present \emph{Cross Entropy-baed Sampling} (CES) 
that conditions on representations learned by $\mathfrak{M}$, 
and approximates the distribution of $S$ through cross-entropy minimization. 

%\textbf{Simple Random Sampling}. 
%Obviously, the most simplely method is simple random sampling.
%For convenience, we denote the test example as $\bm{x}$ and its label as $y$. 
%We use $H(\bm{x})$ represent whether the neural network misclassified for input $\bm{x}$.
%The estimated accuracy through SRS is
%\begin{equation}
%\hat{acc} = \Expect[H(\bm{x})] = \frac{\sum_{i=1}^n H(\bm{x}_i)}{n},
%\end{equation}

%\subsection{Stratification according to confidence}
\subsection{Confidence-based Stratified Sampling}
\label{subsec:condconf}
%As mentioned earlier, one of reasons why conditioning techniques is challenging is that the conditional random variable $Z$ is difficult to form.
%In other words, nearly we cannot find a measurable conditional random variable which has strong correlation to the prediction of example.
As discussed earlier, the key to improve the estimation efficiency is 
to find a random variable $\bm{Z}$ that is strongly correlated to the accuracy of the model $\mathfrak{M}$, 
and whose distribution is easy to sample. 
A natural  choice is the confidence value $c(\bm{x})$ 
provided by some classifier models when predicting the label for $\bm{x}$.
Obviously, predictions with a higher confidence will be more likely to be correct, \emph{if} the classifier is reliable. 

Since the confidence is a bounded scalar, we can divide its range into $k$ sections, 
and stratify the population $S$ into $k$ strata $\{S_1,\cdots,S_k \}$ accordingly. 
Thus the probability of an example belonging to $S_j$ is $P_j=\frac{|S_j|}{|S|}$, $1\leq j\leq k$.
%?? for simplicity
From each stratum $S_j$, randomly taking $n_j$ elements such that $\Sigma_{j}^{k}n_j=n$. 
Then the accuracy of $\mathfrak{M}$ on $S$ is estimated as 

\begin{equation}
\hat{acc} = \sum_{j=1}^k P_j \Expect[H(\bm{x}) \mid \bm{x} \in s_j] = \sum_{j=1}^k P_j \left (\frac{1}{n_j}\sum_{i=1}^{n_j}H(\bm{x}_{j,i}) \right )\,.
\end{equation}

%Elements are randotaking from  with probability $P_j$ such that $\Sigma_{j}^{k}P_j=1$. 
%i.e., randomly taking $n_i = \frac{|S_i|}{|S|}$ examples from $S_i$, $1\leq i\leq k$. 

% method is choosing the confidence of model as conditional variable $Z$,
% i.e. we can simplely divide the population based on confidence of model.
% It indeedly is stratified sampling, for detail, we use the confidence of classification as conditional variable $Z$, and divide the confidence into $k$ stratas $\{Z_1,\dots,Z_k \}$.
% Therefore, we realize the division of population according to the division of confidence, and we can compute the probability $P_j$ of the test example falling at the strata $Z_j$, it equals the proportion of examples in $T$ located in $Z_j$.
% Based on the partition of population by confidence, the estimator can written as follows:

%\begin{equation}
%\hat{acc} = \Expect[\Expect[H(\bm{X}) \mid D] = \sum_{j=1}^k P_j \Expect[H(\bm{\bm{x}}) \mid Z = Z_j],
%\end{equation}

A simple strategy is to use \emph{proportional allocation}, i.e., $n_j={P_j}\cdot{n}$, 
and then the estimation of accuracy becomes $\hat{acc}_{\text{prop}} = \frac{1}{n}\sum{H(\bm{x}_{j,i}})$.
However, although safe~\cite{Chen_2001_JSS}, proportional allocation may be suboptimal. 

The variance of a stratified estimator is the sum of 
variances in each strata: $\Var(\hat{acc})=\Sigma_{j=1}^k{\Var(H(\bm{x}|\bm{x} \in s_j))}$.
% effectiveness of stratified sampling only depends on the estimating $\Expect[H(X|\bm{z_i})]$ for each $\bm{Z_i}$.
% In other words, we need to select more samples in stratas with large variances rather than low variance stratas.
It can be further reduced if we allocate more examples in strata that are less even. 
% Specifically, we can assume that test examples be classified with high confidence are often classified correctly, while low confidence test examples are are more likely to perform differently.
Specifically, we can guess that a stratum with a lower confidence should fluctuate more in the accuracy. 
So we should take more examples from those low-confidence strata. 
%According to this, it is reasonable to label more test examples in low confidence stratas.

The optimal stratification and example allocation depend on 
the actual distribution of the variance of $H(\bm{X})$ conditioned on the confidence, 
which is not known \emph{a priori}. They have to be determined according to experience and pilot experiments. 

%\smallskip
Unfortunately, CSS is not robust for operational DNN testing. 
It performs very well when the model is perfectly trained for the operation context. 
However, as shown in our experiments reported in Section~\ref{sec:Evaluation}, 
when the model is not well-fitted to the operational data set, its performance drops drastically. 
Note that this unfavorable situation is the motivation for operational DNN testing.  
It is not difficult to see the reason -- when the model predicts poorly in the operation context, 
the confidence values it produces cannot be trusted. 
%As shown in our experiments, for some malicious attacks on the model, such as mislabel for training set, or adversarial examples, the information about the confidence of DNN may misleading the example selection.
In addition, confidence values are not readily available in regression tasks. %the confidence stratified sampling could not applied directly to regression tasks.
Therefore, we propose another variance reduction method based on behaviors of neurons.

%\subsection{Conditioning on representation} 
\subsection{Cross Entropy-based Sampling}
\label{subsec:condrepr}
%Another choice which is more robust than conditional variable of confidence stratified sampling is using the output of the last hidden layer in DNN as conditional random variable $\bm{Z}$.
A better choice for the condition random variable $\bm{Z}$ is the output of neurons in the last hidden layer. 
It is often viewed as a learned representation of the training data 
that makes the prediction easier~\cite[p. 6 and p. 518]{Goodfellow-2016-DL}. 
The rationale behind this choice is manifold. First, although not necessarily comprehensible for human, 
the representation is more stable than the prediction when the operation context is drifting. This is supported by well-known transfer learning practices where only the SoftMax layer is retrained for different tasks~\cite{Huang-2013-ICASSP}. Second, the DNN prediction is directly derived from the linear combination of this layer's outputs, and thus it must be highly correlated with the prediction accuracy. 
Finally, the correlation between the neurons in this layer 
is believed to be smaller than those in previous layers~\cite{bengio2013better}, which facilitates the approximation of their joint distribution that will be used in our algorithm. 

For a trained DNN model $\mathfrak{M}$ we consider its last hidden layer $L$ 
consisting of $m$ neurons denoted by $e_i, i= 1,\dots,m$. 
We divide $D_{e_i}$, the output range of each neuron $e_i$, 
into $K$ equal sections $\{D_{e_i,1},\dots,D_{e_i,K}\}$,
and define function $f_{e_i}(\bm{x}) = j$ if the output of $e_i$ for input $\bm{x}$ belongs to $D_{e_i,j}$, $1\leq j \leq K$. 

Hence %the output space of layer $L$ is divided into $k^m$ regions. In other words, 
the conditional variable $\bm{Z}$ is a vector $(\bm{Z}_1,\dots,\bm{Z}_m)$, $\bm{Z}_i \in \{1,\dots,{k}\}$, $1\leq i \leq m$.
Let $S_{z_1,\dots,z_m} = \{\bm{x} \in S \mid f_{e_i}(\bm{x}) = z_i, 1 \leq i \leq m\}$ be a subset of $S$ whose elements are mapped onto $\bm{z} = (z_1,\dots,z_m)$ by the model.
The probability distribution $P_S(\bm{z})$ of $\bm{Z}$ is defined with the operational data set $S$ as 
\begin{equation} 
 P_S({z}_1,\dots,{z}_m) = \frac{|S_{z_1,\dots,z_m}|}{|S|}\,.
 %\frac{|\{\bm{x} \in S \mid f_{e_i}(\bm{x}) = z_i, 1\leq i \leq m\}|}{|S|}
\end{equation}

%taking values in the m-dimensional space $\Pi_{i=1}^m{D_{e_i}}$.

%has $k^m$ possible values in this sense, and we denote them as $\{\bm{Z}_1,\dots,\bm{Z}_{k^m}\}$.

However, considering the high dimensionality of $\bm{Z}$, it is challenging to take a  
%another challenge of DNNs testing is still exists: 
%In high dimensional case, it is difficult to select 
typical sample $T$ from the whole test set $S$ according to $Z$'s distribution, not to mention applying stratified sampling. 
Note that we cannot use artificial examples generated according to $Z$'s distribution 
because we need to evaluate the model's accuracy on real data in the given operation context. 
Now the problem is how to select a small-size sample from a finite population 
which itself is sparsely distributed in a high dimensional space, 
such that the sample is as ``representative'' as possible for the population.  

To this end, 
%Therefore we have to select example from finite sample set.
%In this situation, 
we propose to select a typical%
\footnote{cf. Shannon's concept of typical set~\cite{mackay2003information, ho2010information}.}
 sample $T$ by minimizing the cross entropy~\cite{Goodfellow-2016-DL} between $P_S(\bm{Z})$ and $P_T(\bm{Z})$:
 
\begin{equation} 
\label{equ:crossentropy}
\begin{aligned}
\min_{T \subset S, |T|=n}  CE(T) &=H(P_S,P_T) \\
&= -\sum_{\bm{z}\in\{1,\dots,K\}^m} P_S(\bm{z}) \log P_T(\bm{z})\,,
\end{aligned}
\end{equation}
where
\begin{equation}
P_T({z}_1,\dots,{z}_m) = \frac{|T_{z_1,\dots,z_m}|}{|T|}\,.
% \frac{|\{\bm{x} \in T \mid f_{e_i}(\bm{x}) = z_i, 1\leq i \leq m\}|}{|T|}
\end{equation}

%, where the definition of typicality is proposed in information theory by Claude Shannon.
%For detail, we can calculate the proportion of conditional variable $\bm{Z}$ located in the section $\bm{Z}_j$.
%For the small selected test set $T$ ($n$ examples) and the whole test set $S$ ($N$ examples), we can calculate two different proportions:
%\begin{equation}
%\begin{aligned}
%& P_S(j) = \frac{|\{\bm{x} \in S \mid f(\bm{x}) \in \bm{Z}_j\}|}{|S|}, j=1,\dots,k^m, \\
%& P_T(j) = \frac{|\{\bm{x} \in T \mid f(\bm{x}) \in \bm{Z}_j\}|}{|T|}, j=1,\dots,k^m,
%\end{equation} 
%where $f(\bm{x})$ is the m-dimensional vector of $L$-layer output.
%For convenience, we also denote the output of neuron $e_i$ as $f_{e_i}(\bm{x})$.
%Therefore, we have $f(\bm{x})  = \left( f_{e_1}(\bm{x}),\dots,f_{e_m}(\bm{x}) \right)$.
%In addition, the proportion of examples located in $S$ is actually the probability distribution of $\bm{Z}$.
%
%For given the whole test set $T$, we can select typical sample set $S$ by minimizing the cross entropy:
%\begin{equation}
%
%\end{equation}

In this high-dimensional case, 
the minimization is hard to compute directly, partially because of the sparseness of $S$ and $T$ in $\bm{Z}$'s space.
% minor.
Fortunately, it is observed that a DNN typically reduces the correlation among neurons 
in the last hidden layer~\cite{koller2009probabilistic}, 
thus we can take an approximation by assuming that they are independent of each other in computing the minimization. 
In this case we can minimize $CE(T)$ through minimizing $\overline{CE}(T)$ the average of the cross entropy between $P_S(\bm{Z})$ and $P_T(\bm{Z})$ on each dimension:

%use the product of each dimension of 
%
%$\bm{Z}$ directly to approximate the high dimensional proportion of $\bm{Z}$ .
%In other words, instead of the approximation of high dimensional cross entropy, we use the approximation of the cross entropy in each dimension (i.e. the output of each neuron).
%In this situation, we only calculate the proportion of examples locate in section $D_{e_i}^{(j)}$, i.e.
%\begin{equation} 
%\begin{aligned}
%& P_{T}^{e_i}(j) = \frac{|\{\bm{x} \in T \mid f_{e_i}(\bm{x}) \in D_{e_i}^{(j)} \}|}{|T|}, j=1,\dots,k \\
%& P_{S}^{e_i}(j) = \frac{|\{\bm{x} \in S \mid f_{e_i}(\bm{x}) \in D_{e_i}^{(j)} \}|}{|S|}, j=1,\dots,k
%\end{aligned}
%\end{equation}
%where $f_{e_i}(\bm{x})$ is the output of $e_i$ for input $\bm{x}$.
%
%^Similarly, we use the average cross entropy of each neuron instead of the original joint cross entropy to select typical samples:
\begin{equation} \label{equ:ACE}
\min_{T \subset S, |T|=n}\overline{CE}(T) = -\frac{\sum_{i=1}^m \sum_{z_i=1}^K P_{S}^{e_i}({z_i}) \log P_{T}^{e_i}({z_i})}{m} \,,
\end{equation}
where 
\begin{equation}
\begin{aligned}
P_{S}^{e_i}(z_i) &= \frac{|\{\bm{x} \in S \mid f_{e_i}(\bm{x}) = z_i \}|}{|S|} \,.
%P_{T}^{e_i}(z_i) &= \frac{|\{\bm{x} \in T \mid f_{e_i}(\bm{x}) = z_i \}|}{|T|}.
\end{aligned}
\end{equation}
and $P_{T}^{e_i}(z_i)$ is defined similarly. 

Furthermore, the optimal solution of $CE(T)$ %average cross entropy is $P_{S}^{e_i}(z_i) = P_{T}^{e_i}(z_i),i=1,\dots,m,z_i=1,\dots,K$.
%It is equivalent to to 
is achieved when $P_S(\bm{z}) = P_T(\bm{z}), \bm{z}\in\{1,\dots,K\}^m $~\cite{shore1980axiomatic}. %if the neurons are independent each other.
%Therefore, cross entropy implies a \textbf{Probability Proportional to size Sampling (PPS)} strategy which is an unbiased estimation.
Therefore, the estimator for model accuracy $\Expect[H(\bm{X})]$ is given by:
\begin{equation}
\begin{aligned}
\hat{acc} &= \sum_{\bm{z} \in \{1,\dots,K\}^m} P_T{(\bm{z})} \Expect[H(\bm{x}) \mid \bm{Z}=\bm{z} ] \\ 
%= \sum_{j=1}^{k^m} P_{T}(j)\Expect[H(\bm{x}) \mid \bm{Z} = \bm{Z}_j] \\
&= \sum_{\bm{z} \in \{1,\dots,K\}^m} P_{T}(\bm{z}) \frac{\sum_{\bm{z} \in T_{z_1,\dots,z_m}} H(\bm{x})}{|T_{z_1,\dots,z_m}|} \\
&= \frac{\sum_{\bm{x} \in T}H(\bm{x})}{|T|} = \frac{\sum_{i=1}^n H(\bm{x}_i)}{n} \,.
\end{aligned}
\end{equation}
Note that this estimator is unbiased according to Equation~\ref{equ:expcond}.

% \subsection{Distribution Coverage Criteria for DL Systems}
% According to the previous analysis, for the sample-selection problem, selecting examples which can minimize the cross entropy will obtain more precise estimate about the accuracy of model.

% However, for the whole test set with $N$ unlabeled examples, it is nearly impossible to choose $n$ examples as the optimal of cross entropy.

To solve the optimization problem of Equation~\ref{equ:ACE},
% Therefore, to establish the small test set $T$, 
we propose an algorithm (Algorithm \ref{alg:1}) similar to random walk~\cite{revesz2005random}.
Elaborately, we first randomly select $p$ examples as the initial sample set $T$, 
and repeatedly enlarge the set by a group $Q^*$ of $q$ examples until we exhaust the budget of $n$. 
At each step, $Q^*$ is selected from $\ell$ randomly selected groups, minimizing the cross entropy.  

%Then we put $q$ examples ($q \ll n$) that can minimize cross entropy into $T$ each time until the size of $T$ reaches $n$.
%In addition, for $q$ examples selection, we still have $\binom{q}{N}$ groups for selecting.
%In fact, we don't need to traverse each group to get a completely accurate optimal solution.
%Hence we only select the group which can minimize the cross entropy from random $\ell$ groups $\{Q_1,\dots,Q_{\ell}\}$.

\begin{algorithm}
\caption{Test Input Selection}
\label{alg:1}
\begin{algorithmic}[1]
\REQUIRE Original unlabled test set $S$, DNN $\mathfrak{M}$, the budget $n$ for labeling inputs.
\ENSURE Selected test set $T$ $(|T| = n)$ for labeling.

\STATE Selecting randomly $p$ examples as the initial test set $T$.
\WHILE {$|T| < n$}
    \STATE Randomly select $\ell$ groups of examples, $Q_1,\dots,Q_{\ell}$. 
           Each group contains $\min(q,n-|T|)$ examples.
    \STATE Choose the group that minimizes the cross entropy, i.e., 
\begin{equation}
Q^* = \min_{Q_i} \overline{CE}(T \cup Q_i), i=1,\dots,\ell. 
\end{equation}
    \STATE $T \leftarrow T \cup Q^*$.
\ENDWHILE
\end{algorithmic}
\end{algorithm}

% we need to discuss the connection to variational inference here. 

%\smallskip
Finally, there is an intrinsic connection between structural coverage and the cross entropy in Equation~\ref{equ:crossentropy}. 
Structural coverages actually assume the probability of $P_S(\bm{z})$ to be a constant. 
In this case, minimizing $CE(T)$ becomes maximizing $\sum_{\bm{z}\in\{1,\dots,K\}^m} \log P_T(\bm{z})$,
which equals to maximizing $\prod_{\bm{z}\in\{1,\dots,K\}^m} P_T(\bm{z})$. \linebreak
Since $\sum_{\bm{z}\in\{1,\dots,K\}^m} P_T(\bm{z}) = 1$, it is to even the distribution $P_T(\bm{z})$, 
which is actually to maximize the coverage so that more instances of $\bm{z}$ are covered.

%% file: evaluation.tex
% !TEX root = DistCov.tex

\section{Evaluation}
\label{sec:Evaluation}

Cautious readers may have noted that our approach leverages several heuristics and approximations, 
including the stableness of the representation learned by a DNN model despite of the possible drift of its operational data, 
the independence between the outputs of neurons in the last hidden layer, 
and the optimization through random walk in Algorithm~\ref{alg:1}.
Thus, a systematic empirical evaluation is needed to validate the general efficacy of the approach. 

In the following, we first briefly introduce the implementation of our CSS and CES approaches,  
then discuss some different situations faced by operational DNN testing, 
and how experiments are designed accordingly. 
After that we present the results of the experiments, 
which unanimously confirm that our approach greatly improves testing efficiency. 

\subsection{Implementation}

We implemented our approach using Tensorflow 1.12.0 and Keras 2.2.4 DL frameworks.
The code, along with additional experiment results%
\footnote{Besides those discussed this Section, 
additional experiments were carried out to validate the superiority of the last hidden layer over other layers 
as the learned representation to condition on in CES, 
to explore whether the surprise value~\cite{Kim2019ICSE} can be used as an alternative for the confidence in CSS, and to examine the relative efficiency of CES over SRS with relatively bigger samples. Due to the page limit, we cannot include them in this paper.}, 
can be found at \url{https://github.com/Lizenan1995/DNNOpAcc}.

For CSS, we use an optimal setting achieved through pilot experiments.
The population is partitioned into three strata. 
The 80\% examples of the whole operational dataset with the highest confidence are assigned to the first stratum, 
the next 10\% to the second stratum, 
and the lowest 10\% to the third stratum. To draw a sample with size $n$, we take $n \cdot 20\%$, $n \cdot 40\%$, and $n \cdot 40\%$ examples from the three strata, respectively. 
% 
%
%we rank the examples and divide them into three stratas according to the confidence of the model.
%The proportion of examples located in each strata is 10\%,10\% and 80\%.
%In first strata, i.e. examples with the lowest confidence, we randomly select $n \cdot 40\%$ from them.
%Similarly, in the second strata we also randomly select examples.
%At the last strata, i.e. from the high confidence examples, we randomly selected only $n \cdot 20\%$ of them.
%Therefore, the estimate of accuracy is
%\begin{equation}
%\hat{acc} = 0.1 E[H(\bm{x}) \mid Z_1] + 0.1 E[H(\bm{x}) \mid Z_2] + 0.8 E[H(\bm{x}) \mid Z_3],
%\end{equation}
%where $E[H(\bm{x}) \mid Z_i], i=1,2,3$ is the average accuracy of examples located in strata $Z_i$.

For the CES approach that conditions on representation, 
we set $K$ the number of sections for each neuron to $20$. This is not necessarily the best number, 
but is reasonable considering the tens-to-few-hundreds examples are expected to be sampled.
In implementing Algorithm~\ref{alg:1}, we select $p=30$ initial examples, and enlarge the set by $q=5$ examples in each step. 
The number of random groups  examined in each step $\ell$ is set to $300$.
These parameters are fixed in all experiments except for those with very small operational test sets (to be detailed in Section~\ref{subsubsec:smallopset}). 
Further optimizations of these parameters is possible, 
but the above values are already sufficient for achieving a significant efficiency improvement over SRS.

\subsection{Experiment Design}

Generally, operational DNN testing is to detect the performance loss of a DNN model when used in a specific operation context. Here we assume that the model is well trained with its training set, 
and do not explicitly consider the problems usually addressed in the training process, such as under-fitting or over-fitting. 
Hence the performance loss is likely to be caused by   
\begin{itemize}
	\item \emph{Polluted training set}. The training set is mutated by an accident or malicious attack, 
			and thus a mutated model is generated.   
	\item \emph{Different system settings}. For example, the model might be trained with high-resolution examples 
			but used to classify low-resolution images due to the limitation of the camera equipped in the system. 
	\item \emph{Different physical environment}. For example, the lightening condition may vary in the operation environment.  
\end{itemize}

In addition, we need to consider the differences in the purpose (classification or regression), 
the scale of DNN models, and the size of unlabeled operational test sets. 

With these considerations, as shown in Table~\ref{tab:ExpDesign}, we designed 20 experiments in total, 
which varied in the training sets, the DNN models, the operational testing sets and thus the actual operational accuracies.
DNN models with very different structures were used in these experiments.  
Table~\ref{tab:layersneurons} lists the numbers of their layers and neurons. 

\begin{table}[htb]
\caption{Experiment settings and E-Value results}
\centering
\setlength{\tabcolsep}{.36em}
\begin{threeparttable}
\begin{tabular}{|p{0.4cm}<{\centering}|p{1.2cm}<{\centering}|c|p{1.6cm}<{\centering}|c|p{1.2cm}<{\centering}|}
\hline
\multirow{2}*{\textbf{No.}} & \textbf{Train} & \multirow{2}*{\textbf{Model}} & \textbf{Operational} & \textbf{Actual} & \textbf{E-Value} \\
& \textbf{Set} &  & \textbf{Test Set} & \textbf{ Acc. (\%)} & \textbf{{\footnotesize CES/SRS} } \\
\hhline{|======|}
1 &  \multirow{3}*{MNIST}  & LeNet-1 & \multirow{6}*{MNIST} & 93.1 & 0.588 \\
2 & & LeNet-4 & & 96.8 & 0.655 \\
\cline{3-3}
3 & & \multirow{4}*{LeNet-5} & & 98.7 & 0.708 \\
\cline{2-2}
4 & Mutant1$^{a}$ &  & & 79.5 & 0.499 \\
5 & Mutant2$^{a}$ &  & & 77.3 & 0.380 \\
6 & Mutant3$^{a}$ &  & & 79.1 & 0.478 \\
\hhline{|======|}%\hline
7 & \multirow{6}*{Driving} & Dave-orig & \multirow{2}*{Driving} & 90.4$^{b}$ & 0.592 \\
8 & & Dave-drop & & 91.8$^{b}$ & 0.588 \\
\cline{4-4}
9 & & Dave-orig & \multirow{2}*{patch} & 88.3$^{b}$ & 0.426 \\
10 & & Dave-drop & & 83.5$^{b}$ & 0.526 \\
\cline{4-4}
11 & & Dave-orig & \multirow{2}*{light} & 89.8$^{b}$ & 0.375 \\
12 & & Dave-drop &  & 88.5$^{b}$ & 0.481 \\
\hhline{|======|}%\hline
13 & \multirow{4}*{ImageNet} & VGG-19 & \multirow{2}*{ImageNet} & 72.7 & 0.567 \\
14 & & ResNet-50 & & 75.9 & 0.471 \\
\cline{4-4}
15 & & VGG-19 & \multirow{2}*{resolution} & 63.3 & 0.470 \\
16 & & ResNet-50 & & 68.8 & 0.436 \\
\hhline{|======|}%\hline
%17 & \multirow{2}*{Mutant1$^a$} & \multirow{2}*{LeNet-5} & MNIST-100$^{c}$ & $78.6 \pm 4.3$  & $44.3 \pm 15.4$ \\
%18 &                                         &                                   & MNIST-300$^{c}$ & $78.4 \pm 2.1$  & $37.5 \pm 11.5$ \\
%\cline{2-4}
%19 & \multirow{2}*{Driving}         & \multirow{2}*{Dave-orig} & patch-100$^{c}$ & $86.9 \pm 1.4$  &            $59.4 \pm 26.5$ \\
%20 &                                         &                                     & patch-300$^{c}$ & $88.3 \pm 1.2$   & $54.9 \pm 16.2$ \\
17 & \multirow{2}*{Mutant1$^a$} & \multirow{2}*{LeNet-5} & MNIST-100 & $78.6^{c} $  & $0.443^{c}$ \\
18 &                                         &           & MNIST-300 & $78.4^{c} $  & $0.375^{c}$  \\
\cline{2-4}
19 & \multirow{2}*{Driving}         & \multirow{2}*{Dave-orig} & patch-100 & $86.9^{c} $  &  $0.594^{c} $ \\
20 &                                         &                 & patch-300 & $88.3^{c} $   & $0.549^{c}$ \\
\hline
\end{tabular}
 \begin{tablenotes}
        \footnotesize
        \item[a] The three mutated models are trained by changing the labels of training data: $8\leftrightarrow 0$, $7\leftrightarrow 1$, $9\leftrightarrow 3$, respectively.
        \item[b] Since the steering angle is a continuous value, we use 1-MSE (Mean Squared Error) as the accuracy.
%        \item[c] The experiment was repeated 30 times with randomly selected test sets, thus we present the mean and standard deviation of accuracy and E value as $\mu \pm \sigma$.
        \item[c] It is the mean value of 30 experiments with different randomly selected test sets. 
      \end{tablenotes}
 \end{threeparttable}
\label{tab:ExpDesign}
\end{table}

\begin{table}[htb]
\caption{Layers and neurons of DNN models}
\centering
\setlength{\tabcolsep}{.42em}
\begin{tabular}{|c|c|c|c|c|c|c|c|}
\hline
\multirow{2}*{\textbf{Model}} & \multicolumn{3}{c|}{LeNet-} & \multicolumn{2}{c|}{DAVE-} & \multirow{2}*{VGG-19} & \multirow{2}*{ResNet-50} \\
% \cline{2-6}
& 1 & 4 & 5 & orig & drop & & \\
\hline
\textbf{Neurons} & 52 & 148 & 268 & 1,560 & 844 & 16,168 & 94,059 \\
\hline
\textbf{Layers} &7 & 8 & 9 & 13 & 15 & 25 & 176  \\
\hline
\end{tabular}
\label{tab:layersneurons}
\end{table}

The first group of experiments (No.1-6, results to be discussed in Section~\ref{subsubsec:MNIST} ) were designed to study the effect of a polluted training set, 
and to see whether the conditioning approaches were robust when the actual accuracy varied. 
The second group of experiments (No.7-12, Section~\ref{subsubsec:Driving}) simulated different physical environment conditions, 
and the third group (No.13-16, Section~\ref{subsubsec:ImageNet})  were for different system settings.

\begin{figure*}[ht] 
\centering 
\subfloat[LeNet-1]{
\includegraphics[width=.32\textwidth]{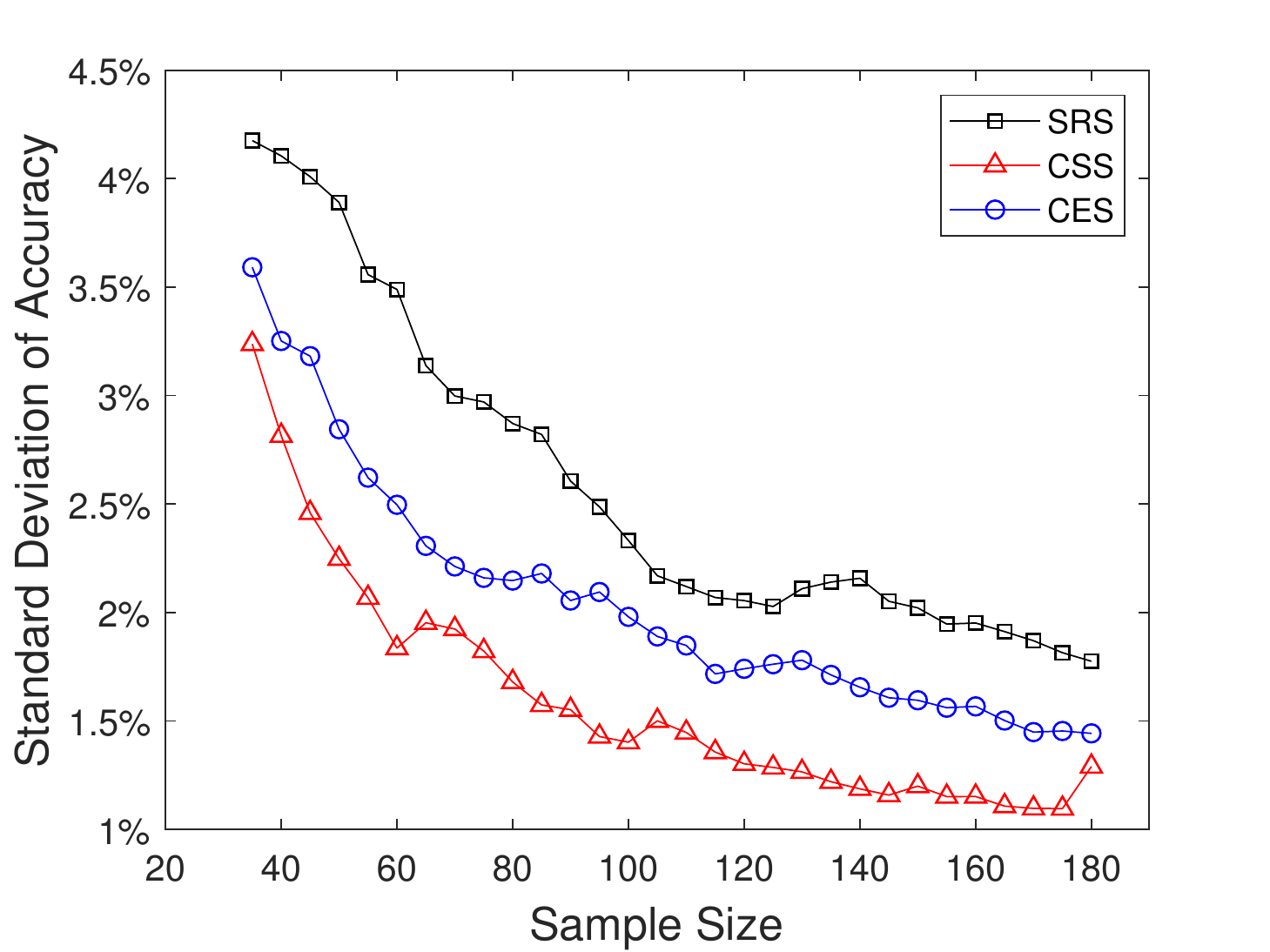}}
\subfloat[LeNet-4]{
\includegraphics[width=.32\textwidth]{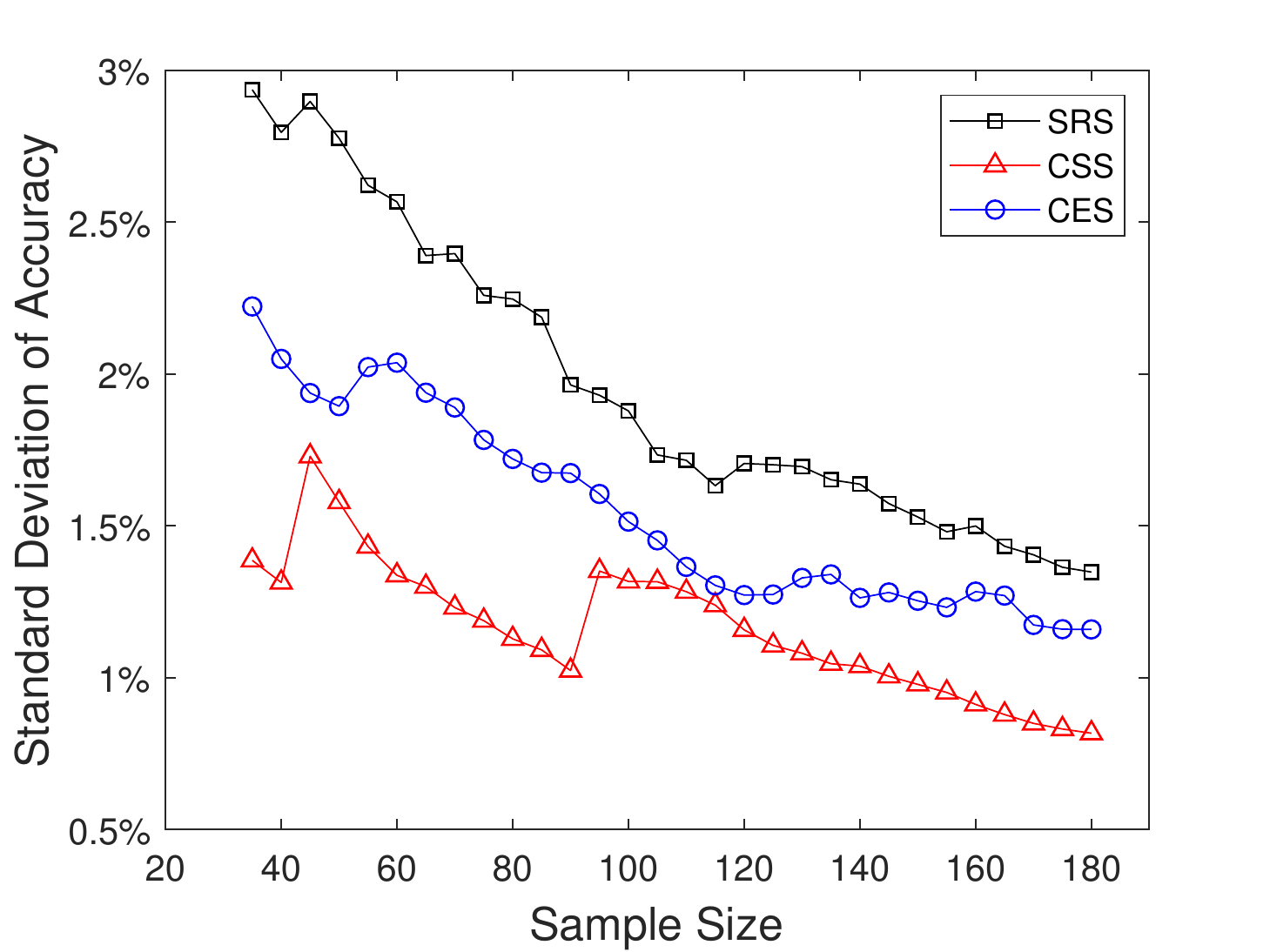}}
\subfloat[LeNet-5]{
\includegraphics[width=.32\textwidth]{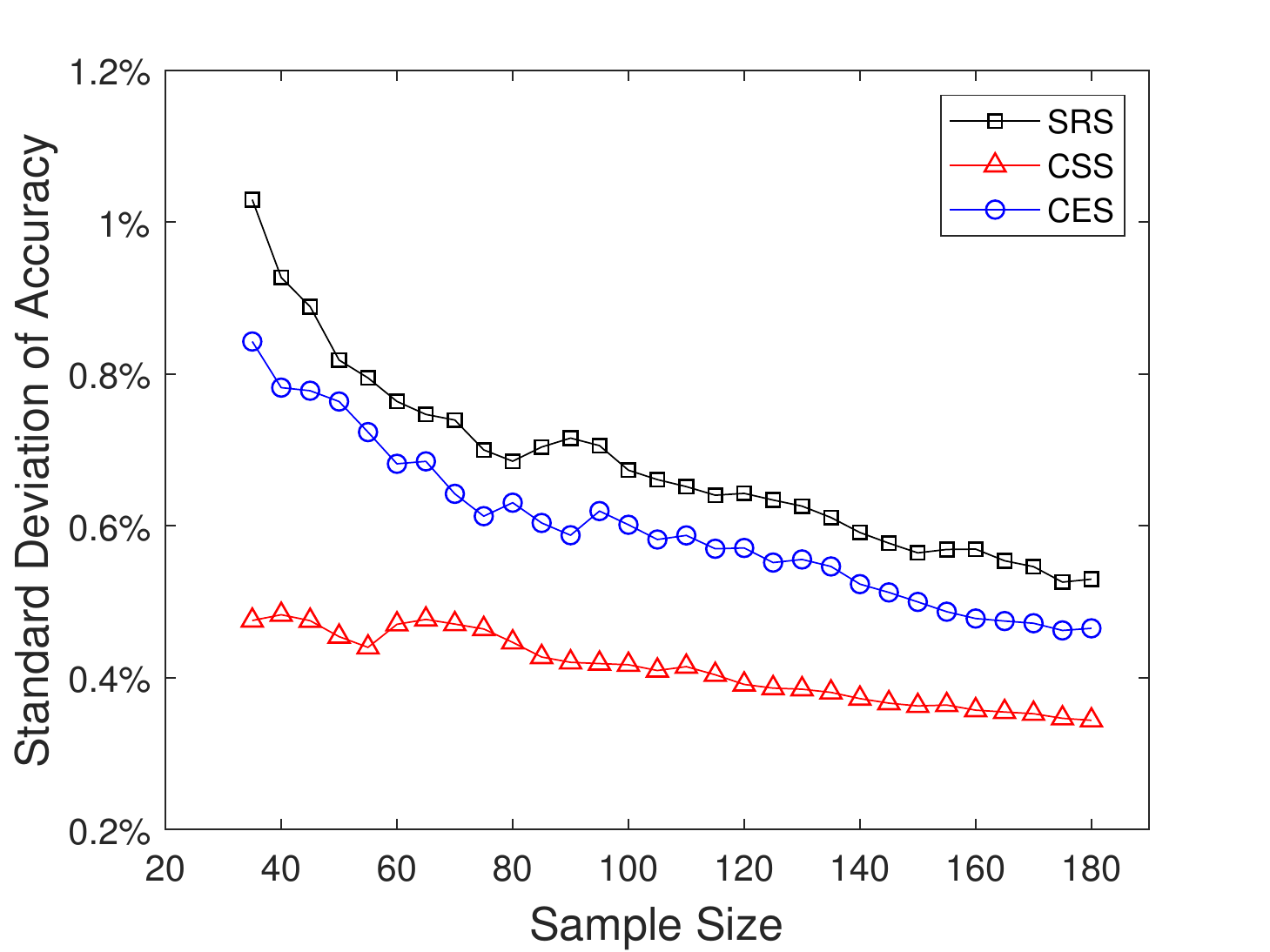}} \\
\subfloat[Mutant-1]{
\includegraphics[width=.32\textwidth]{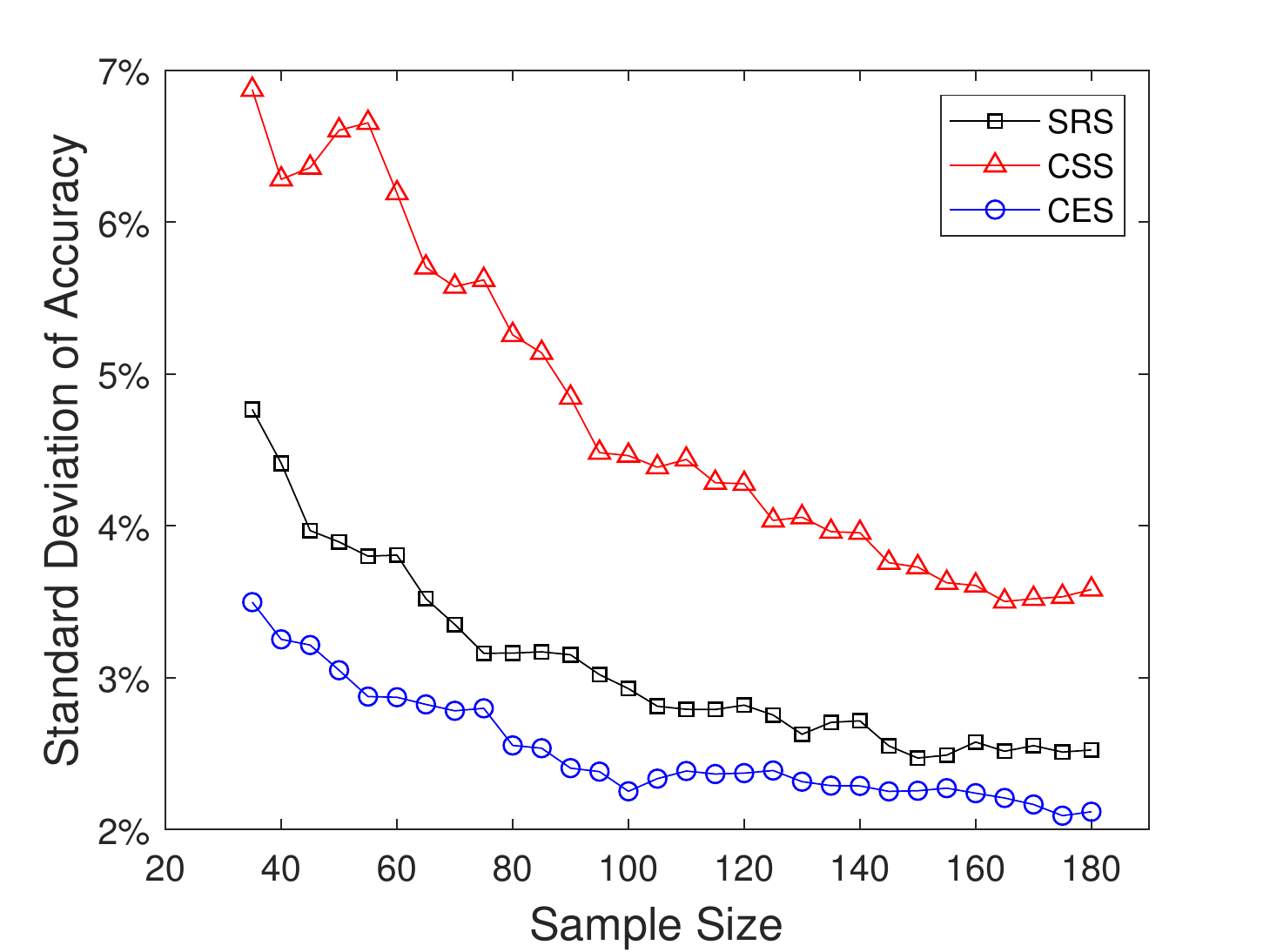}}
\subfloat[Mutant-2]{
\includegraphics[width=.32\textwidth]{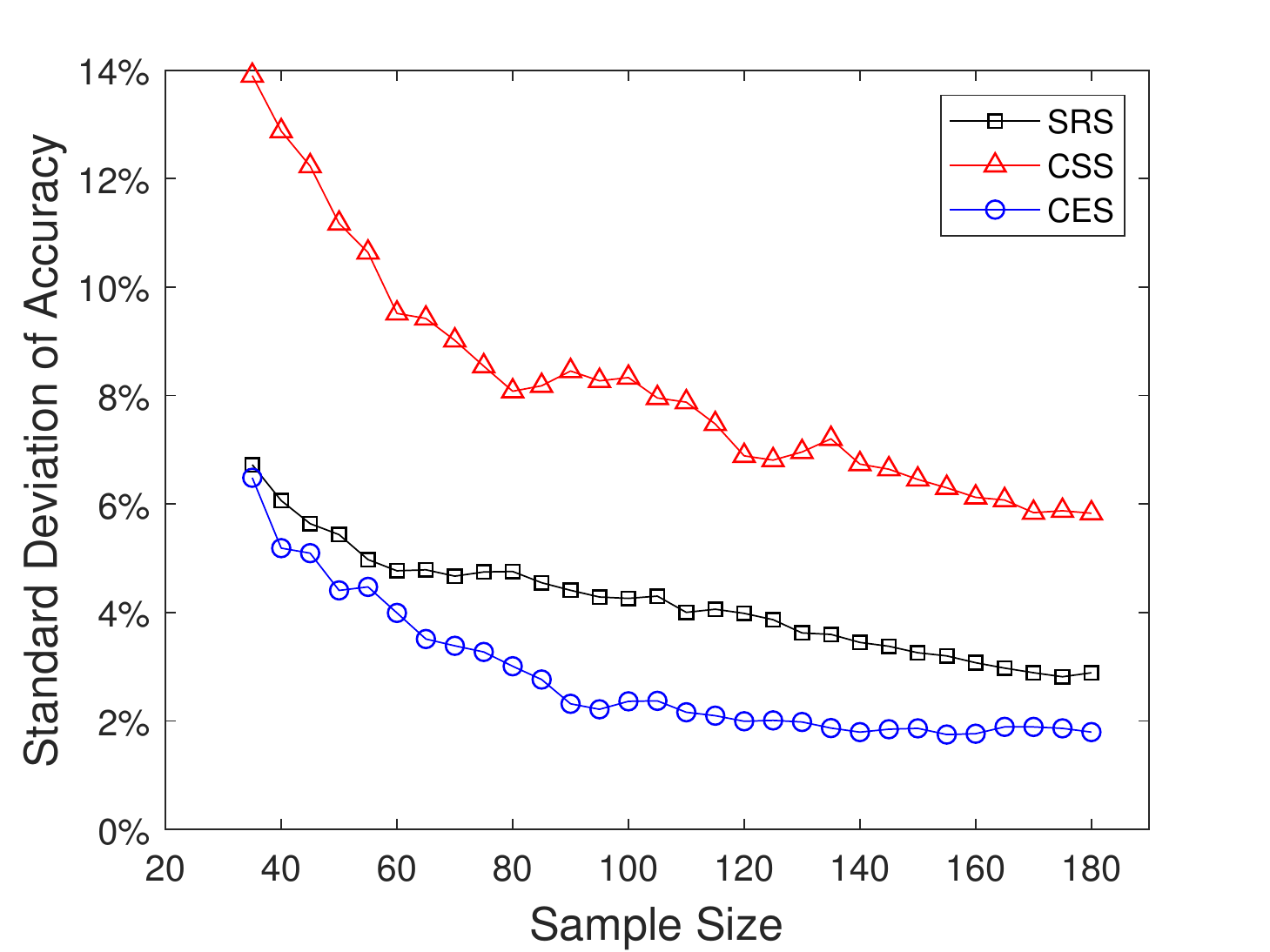}}
\subfloat[Mutant-3]{
\includegraphics[width=.32\textwidth]{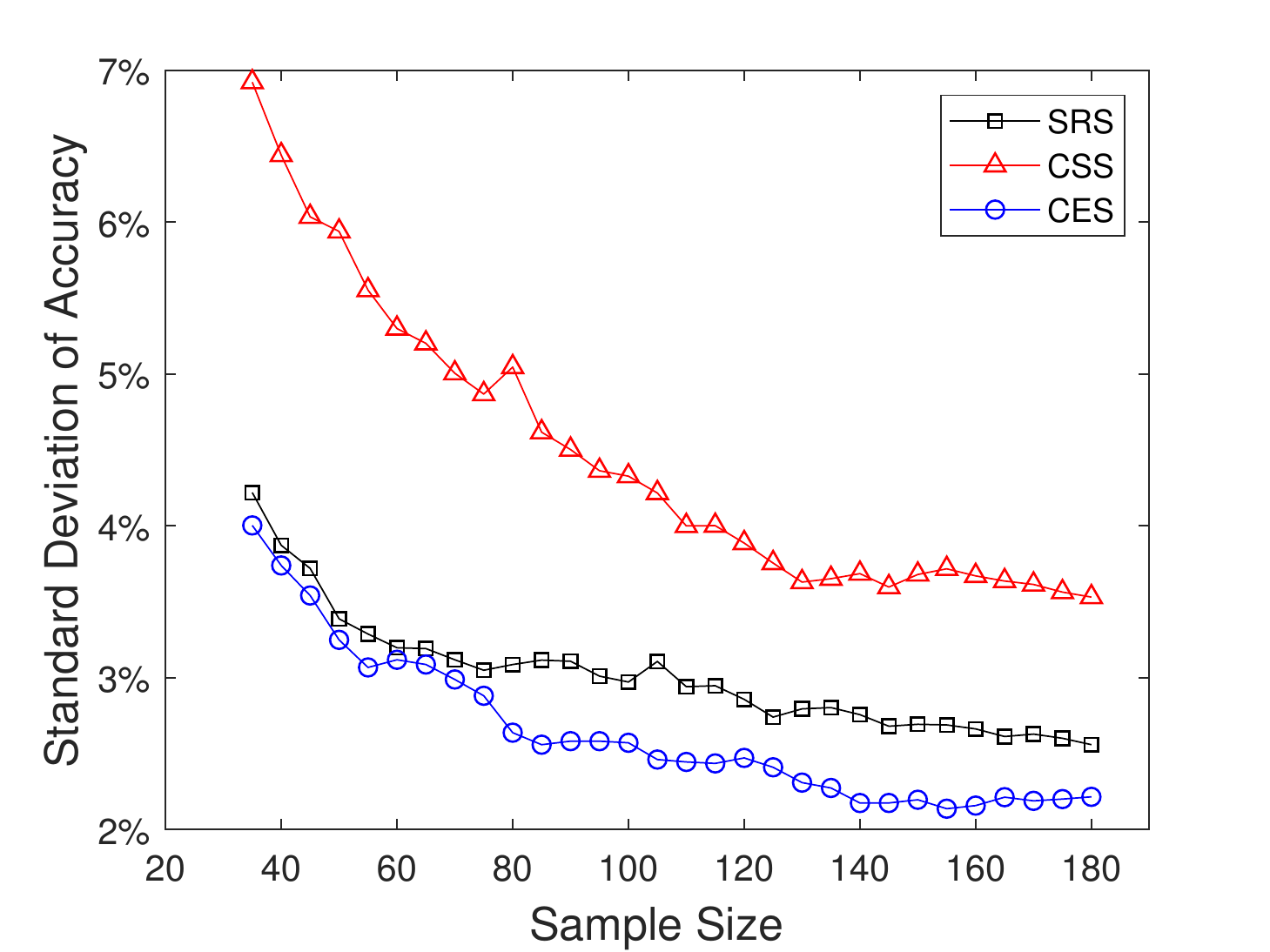}}
\caption{Results of experiments with MNIST}     
\label{fig:mnist}
\end{figure*}

\begin{figure*}[ht] 
\centering 
\subfloat[Original dataset, Dave-orig]{
\includegraphics[width=.32\textwidth]{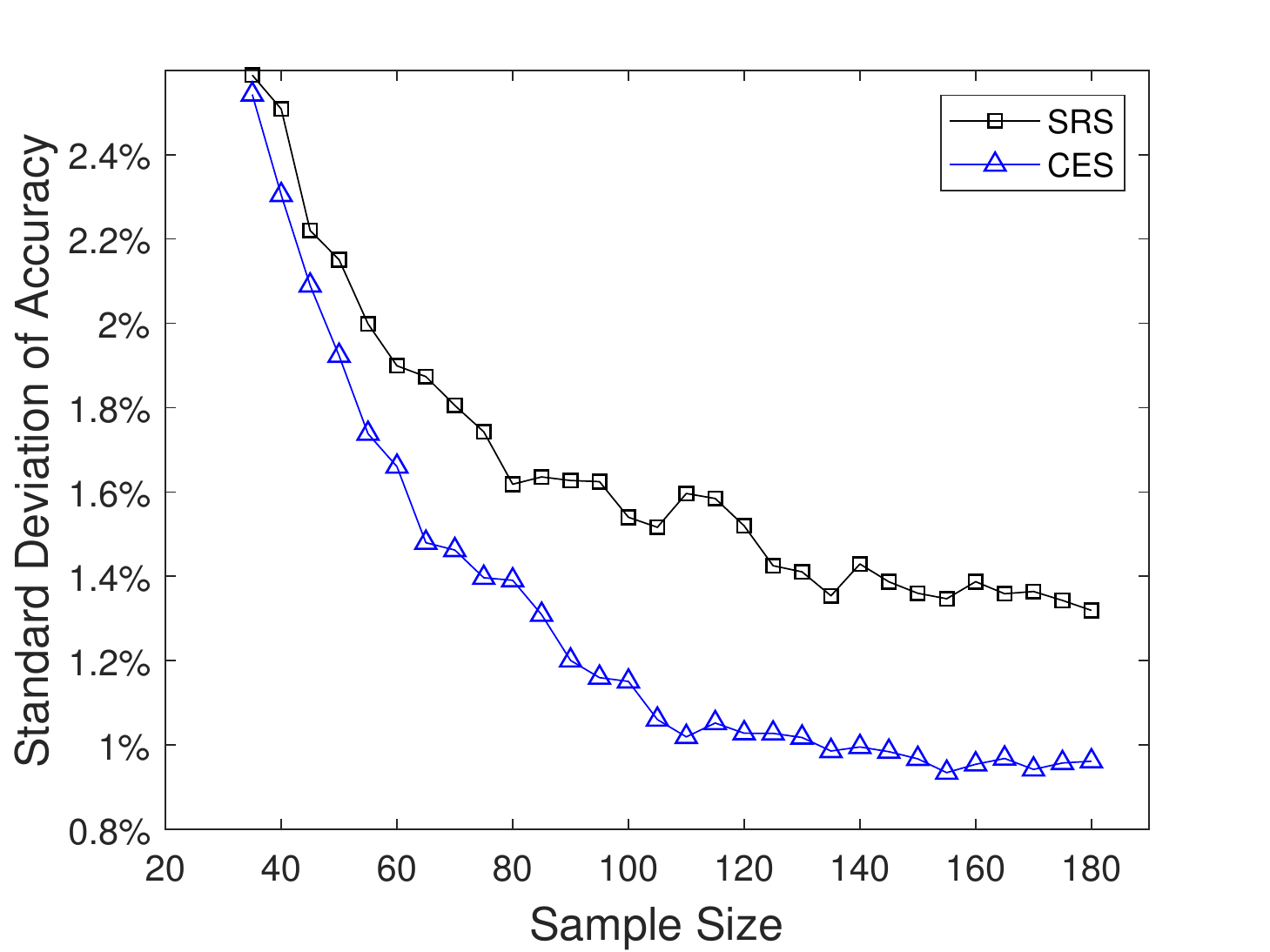}}
\subfloat[Parts blocked, Dave-orig]{
\includegraphics[width=.32\textwidth]{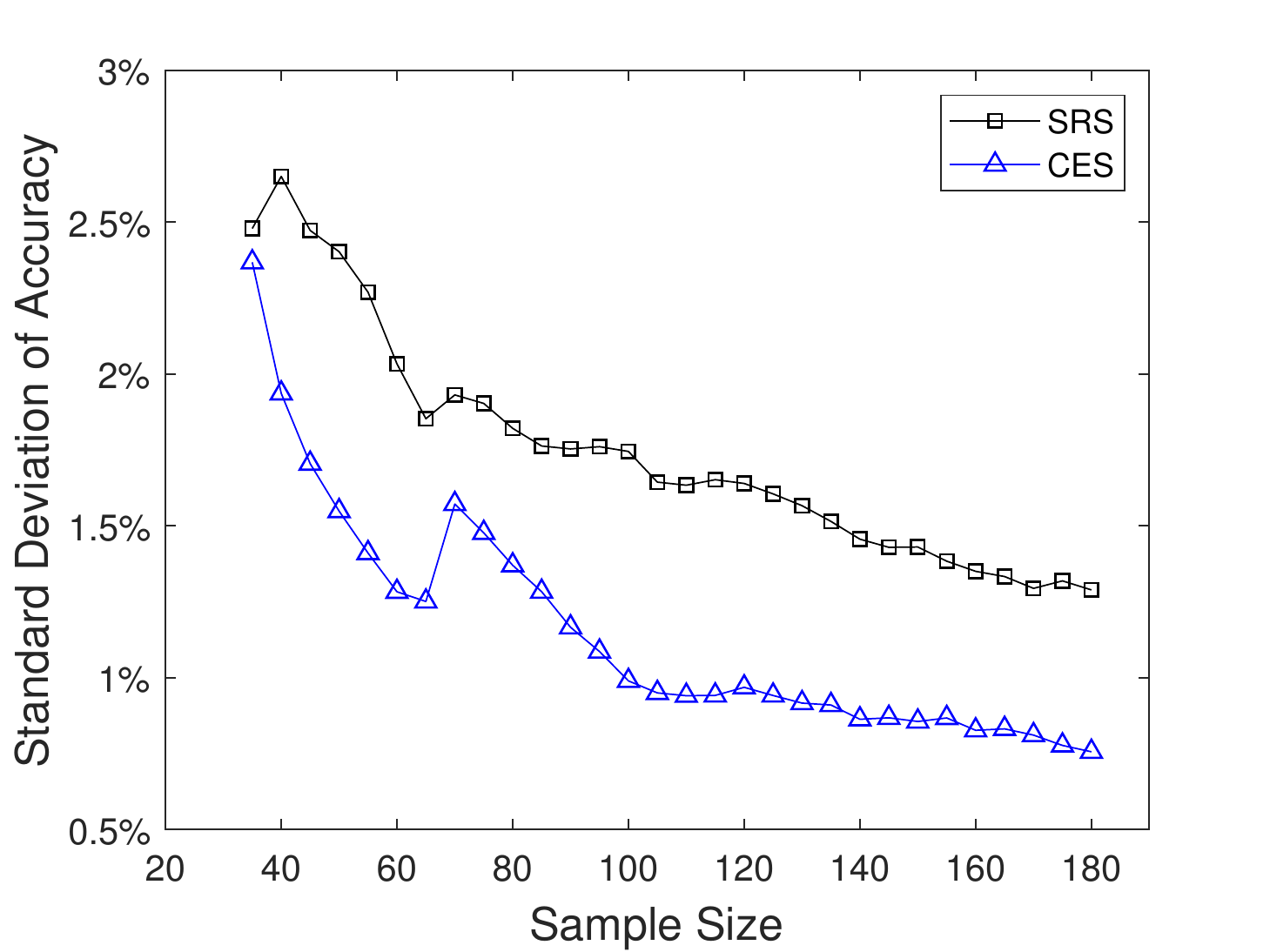}} 
\subfloat[Light modified, Dave-orig]{
\includegraphics[width=.32\textwidth]{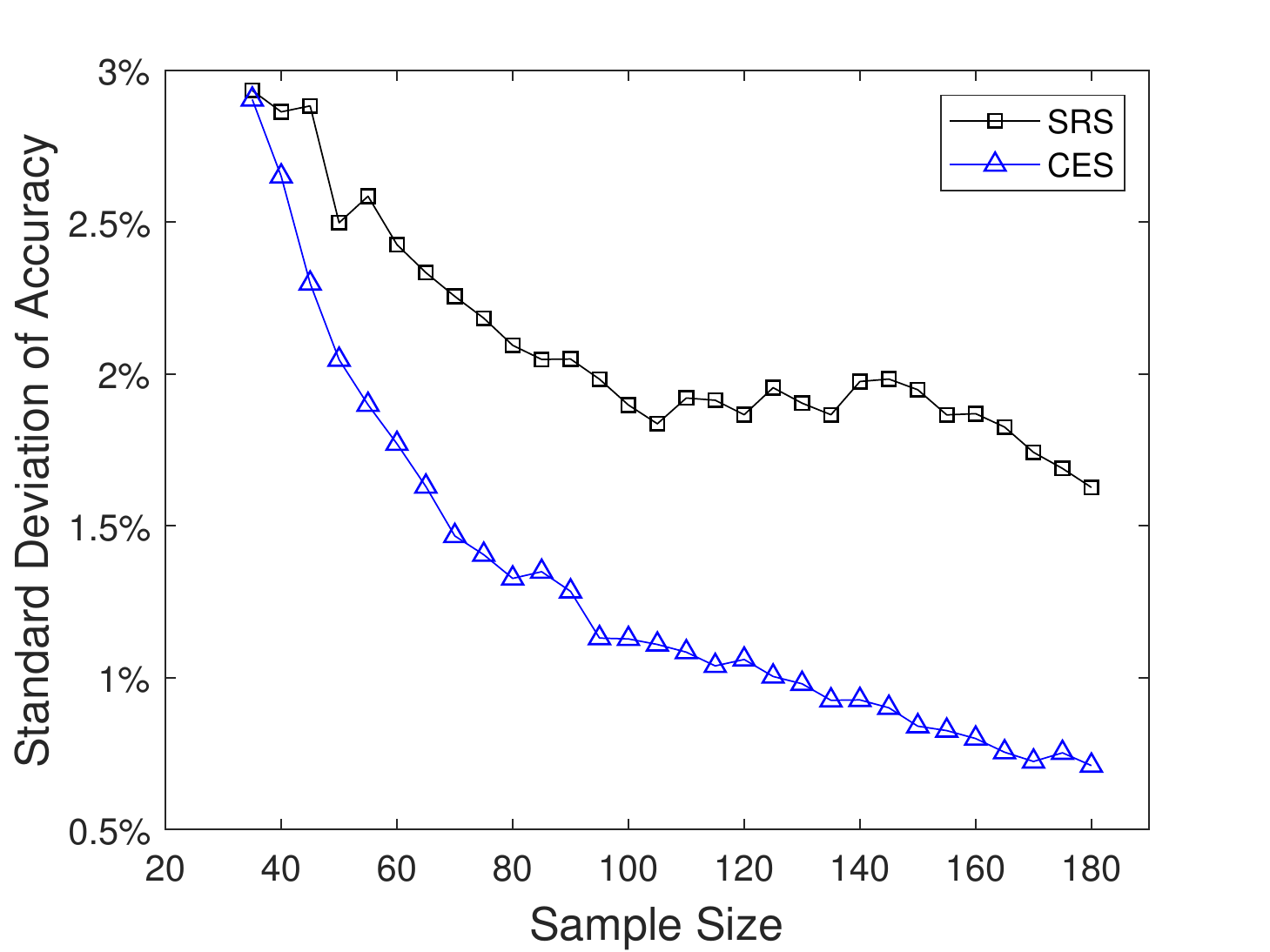}} \\
\subfloat[Original dataset, Dave-drop]{
\includegraphics[width=.32\textwidth]{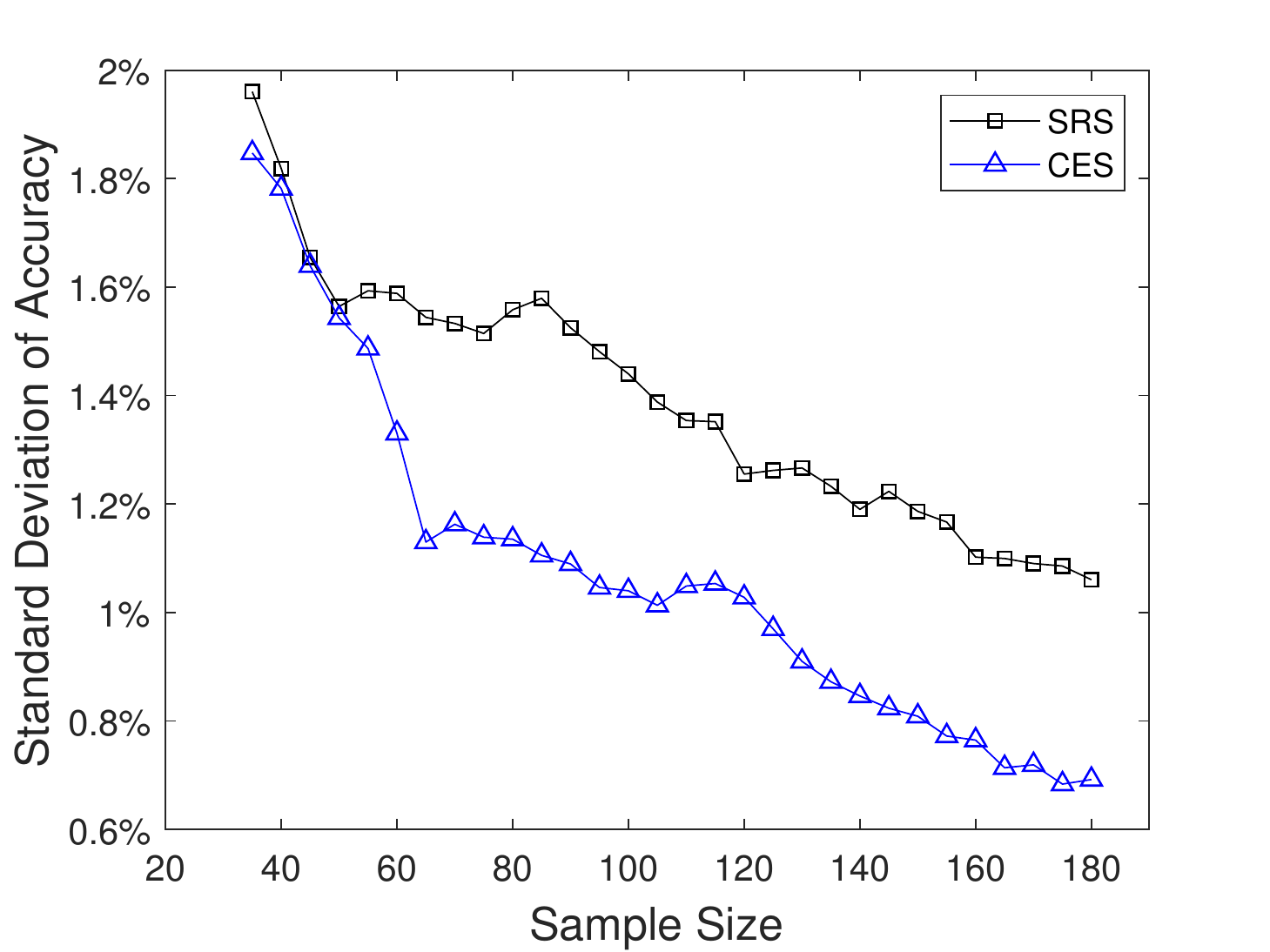}} 
\subfloat[Parts blocked, Dave-drop]{
\includegraphics[width=.32\textwidth]{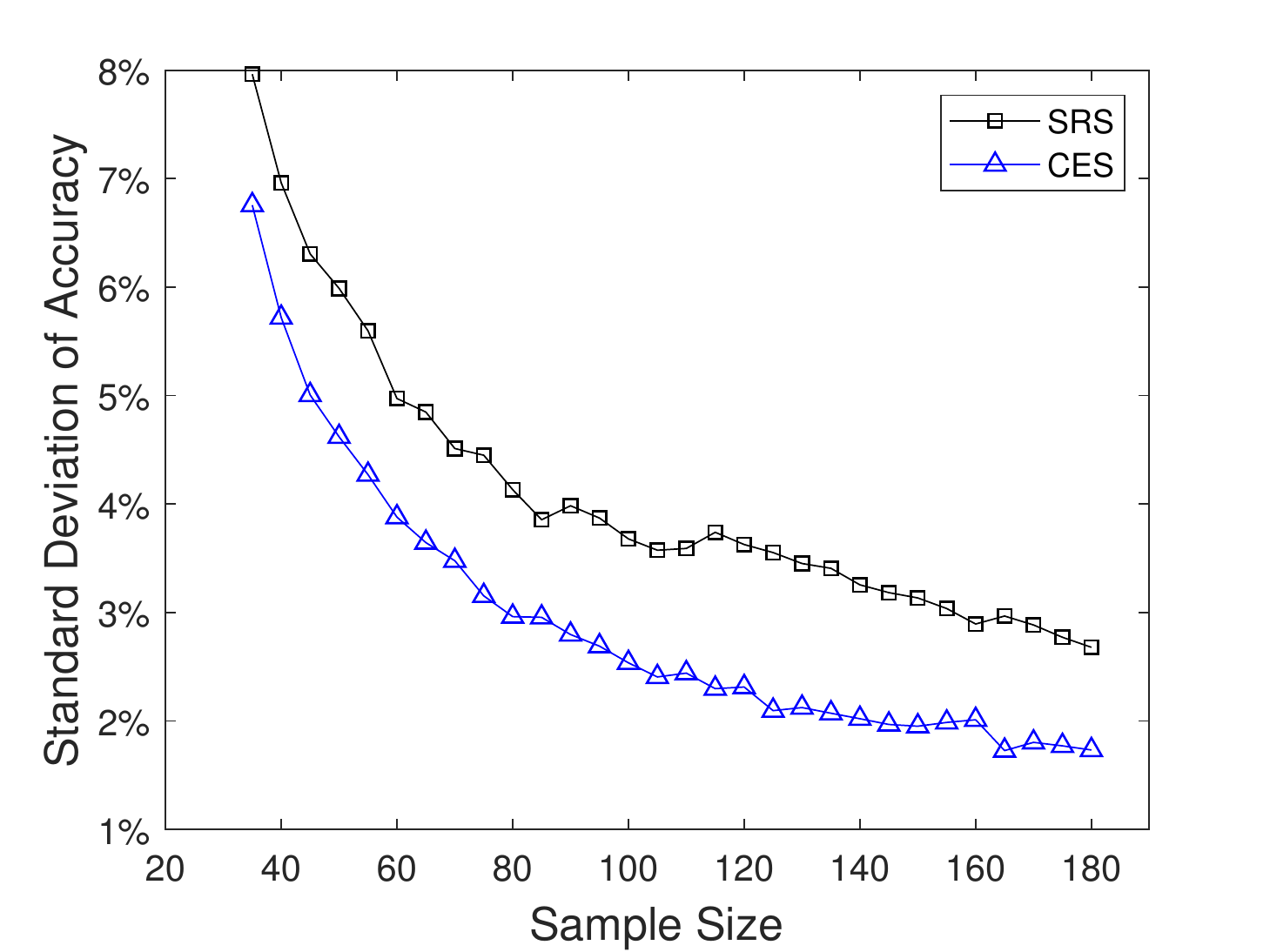}} 
\subfloat[Light modified, Dave-drop]{
\includegraphics[width=.32\textwidth]{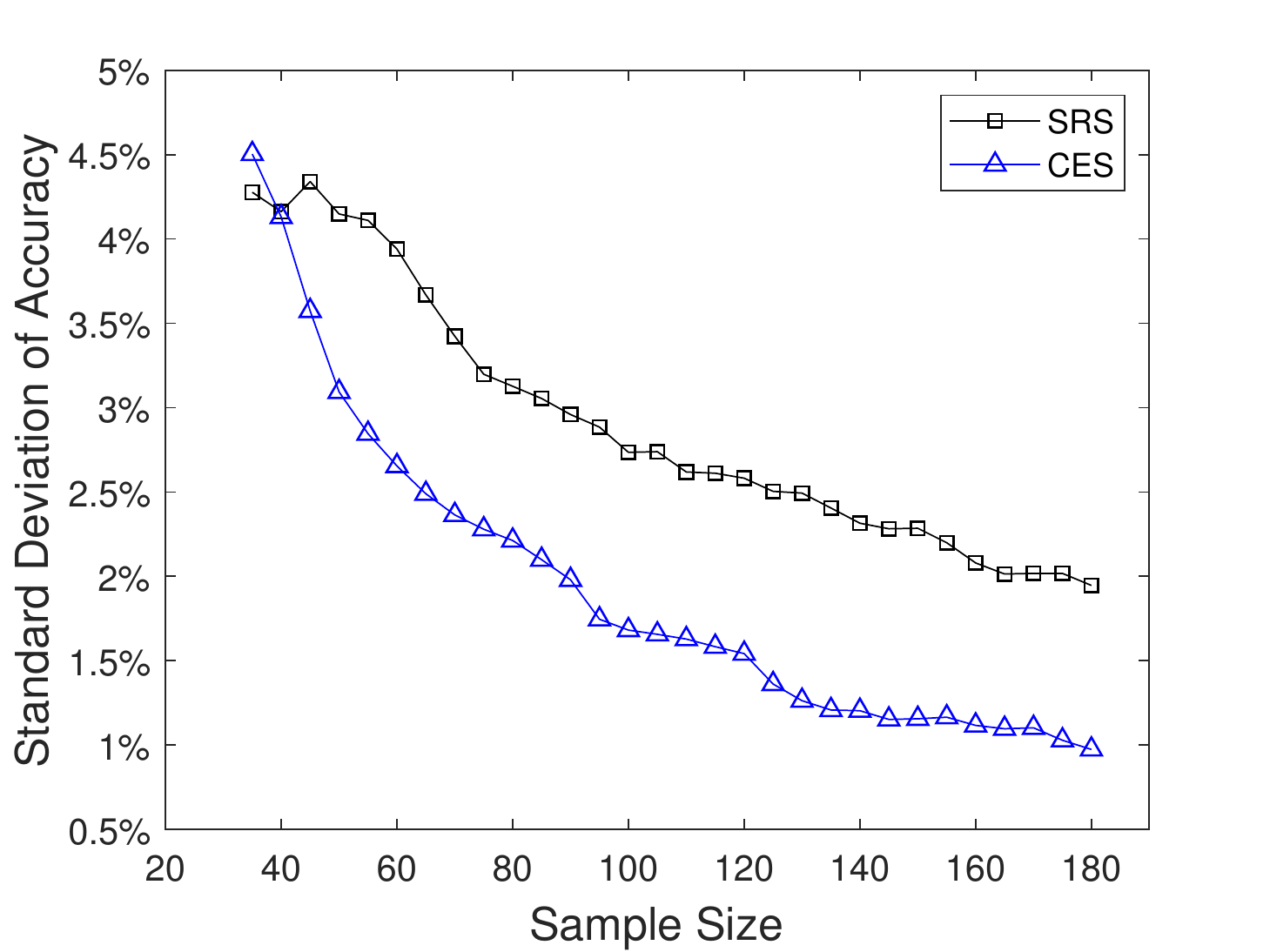}}
\caption{Results of experiments with Driving dataset}     
\label{fig:driving}
\end{figure*}

In these experiments, 
we compared the mean squared errors $MSE(\hat{acc})$ of 
different estimated accuracy $\hat{acc}$ from different estimators. 
They were the SRS estimator (Section~\ref{subsec:conditioning}), 
the CSS estimator (Section~\ref{subsec:condconf}), 
and the CES estimator (Section~\ref{subsec:condrepr}). 
Each experiment was repeated 50 times on each sample size of $35, 40,\dots,180$.%
\footnote{Focusing on estimation with small size samples, 
here we only give results up to sample size 180. 
However, an additional experiment presented at our code website demonstrated  
that the relative efficiency of CES over SRS is quite stable when the sample size grew up to 2500. }
The mean square error was computed as $\frac{1}{50}\sum_{i=1}^{50}(\hat{acc_i}-acc)^2$, 
where $\hat{acc_i}$ and $acc$ were the estimated and actual operational accuracy, respectively. 
Note that because all these estimators are unbiased, 
the $MSE$ can be regarded as the estimation variance, 
whose square root, i.e., the standard deviation, is plotted in the Figures~\ref{fig:mnist}, \ref{fig:driving}, and~\ref{fig:imagenet}.

%\vspace*{\floatsep}% https://tex.stackexchange.com/q/26521/5764

The final group of experiments (No. 17-20, Section~\ref{subsubsec:smallopset}) were designed to 
see whether our approach still worked in cases that 
only a small number of unlabeled operational examples were available. 
We tested cases of taking a sample of size 30 from an operational test set of size 100, 
and of taking a sample of 50 from 300. %for each of the Mutant1/LeNet-5 and Driving/Dave-orig configurations. 

In addition to the visual plotting of standard deviations, we also computed the 
relative efficiency of two estimators as 
$E = \sigma_1^2/\sigma_2^2$ the ratio between their variances. 
Considering that the variance of an SRS estimator is inversely proportional to the sample size, 
$E$ actually indicates the rate of sample size reduction. 
We list the averaged $E$ values (the smaller the better) 
of CES to SRS for each experiment in the last column of Table~\ref{tab:ExpDesign}.

Experiments with the ImageNet dataset (No. 13-16) were conducted on a Linux server with two 10-core 2.20GHz Xeon E5-2630 CPUs, 124 GB RAM, and 2 NVIDIA GTX 1080Ti GPUs, and other experiments were on a Linux laptop with an 2.20GHz i7-8750H CPU, 16 GB RAM, and a NVIDIA GTX 1050Ti GPU. 
For a feeling about the computational cost of sampling with CES, 
we observed that, to select out a sample of size 100, 
it took 8.27s for MNIST/LeNet-5, 20.50s for Driving/Dave-orig, 
and 420.09s for ImageNet/VGG-19.  

\subsection{Experiment Results}

\subsubsection{Experiments with the MNIST dataset}
\label{subsubsec:MNIST}
Experiments 1-6 were conducted on the MNIST dataset~\cite{lecun1998gradient} that is widely used in machine learning research. 
It is a handwritten digit dataset consisting of 60,000 28$\times$28-pixel training images and 10,000 testing images in 10 classes. 
With this dataset, we trained three LeNet family models (LeNet-1, LeNet-4, and LeNet-5)~\cite{lecun1998gradient}. % for analysis.  

Experiments 1-3 tested the ideal situation where the training set and testing set were both original.
Experiments 4-6 the models were trained with mutated training set,
but tested with the original testing images as the operational dataset.
The mutated training set is obtained by exchanging the labels of training data.

From the first row of Figure~\ref{fig:mnist}
 we can see that, 
when there was no divergence between the training data and the operational data,  
the CSS estimator performed particularly well, achieving a $0.379$, $0.372$, and $0.198$ 
average efficiency relative to SRS for the three models tested, respectively. 
However, the second row of Figure \ref{fig:mnist} tells a completely different story. 
In this case, the training set had been mutated, 
and CSS performed very bad, with a $3.263$, $3.836$, and $2.919$ average relative efficiency, respectively.
Clearly, the CSS estimator is not robust to the divergence between the training data and operational data. 

On the contrary, the CES estimator consistently outperformed SRS in both cases. 
From Figure~\ref{fig:mnist} and the relative efficiency values listed in Table~\ref{tab:ExpDesign}, 
we can see that CES only required about a half of labeled data to achieve the same level of precision 
of estimation.

The result of Experiment 3 also suggests that, when the operational accuracy was very high (about 99\%), 
the benefit of our CES diminished, to a level saving about 30\% labeling effort. 
However, this should not be a problem because the accuracy is high, and in this situation we can switch to CSS if needed.

\subsubsection{Experiments with the Driving dataset}
\label{subsubsec:Driving}
The next 6 experiments were conducted on the Driving dataset%
\footnote{\url{https://udacity.com/self-driving-car}}.
It is the Udacity self-driving car challenge dataset containing 101,396 training and 5,614 testing examples.
This is a regression task that predicts the steering wheel angle 
based on the images captured by a camera equipped on a driving car. 
Two pre-trained DNNs (DAVE-orig and DAVE-dropout~\cite{bojarski2016ArXiv}) were used in our experiments. 

%
%\begin{figure}[!ht]
%\centering
%\subfloat{
%\includegraphics[width=.15\textwidth]{figures/driving.png}}
%\subfloat{
%\includegraphics[width=.15\textwidth]{figures/light1.png}} 
%\subfloat{
%\includegraphics[width=.15\textwidth]{figures/light2.png}} 
%\\
%\subfloat{
%\includegraphics[width=.15\textwidth]{figures/driving.png}}
%\subfloat{
%\includegraphics[width=.15\textwidth]{figures/black1.png}} 
%\subfloat{
%\includegraphics[width=.15\textwidth]{figures/black2.png}} 
%\\
%\caption{Mutations of the Driving test data}
%\label{fig:patchandlighting}
%\end{figure}

\begin{figure}[!ht]
\centering
\subfloat[Original]{
\includegraphics[width=.15\textwidth]{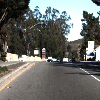}}
\subfloat[Darkened]{
\includegraphics[width=.15\textwidth]{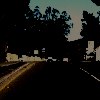}} 
\subfloat[Blocked ]{
\includegraphics[width=.15\textwidth]{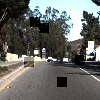}} 
\caption{Mutations of the Driving test data}
\label{fig:patchandlighting}
\end{figure}

%based on the DAVE-2 self-driving car architecture for evaluation.
Experiments 7 and 8 tested the ideal situation where the operational test set is untouched. 
Experiments 9-12 used mutated operational data simulating unideal physical environment conditions, 
with the two methods used by DeepXplore~\cite{Pei_2017_SOSP}: 
occlusion by small rectangles simulating an attacker blocking some parts of a camera (Experiments 9 and 10) 
and lighting effects for simulating different intensities of lights (Experiments 11 and 12). 
Figure \ref{fig:patchandlighting} illustrates the mutations.   
%
%Experiment 9 and 10 used mutated operational data 

Figure~\ref{fig:driving} plots the mean squared errors of CES and SRS for each of the 6 experiments. 
Because the regression models did not provide confidence values, the CSS estimator could not be applied in these experiments. 

%
%includes the detailed MSE for Driving dataset, each column represents original dataset, operational dataset modified by adding black patch and operational dataset modified by changing light, respectively.
%And the first row is experment based on DAVE-orig and the second is based on DAVE-dropout.
The results show that  CES  also worked well for regression tasks, 
%with or without the divergence between the training data and operational testing data. 
and achieved 0.375-0.526 relative efficiency w.r.t.\ SRS when divergence between training data and operational data existed. 

%According to the $E$ values in Table \ref{tab:1}, the CES nearly can save half sample size comparing with the SRS.

\subsubsection{Experiments with the ImageNet dataset}
\label{subsubsec:ImageNet}
The ImageNet~\cite{imagenet_cvpr09} dataset is chosen for the last 4 experiments.
It is a large collection of more than 1.4 million images as the training data and other 50,000 as the test data, 
in 1,000 classes. 
Two large scale pre-trained models, viz.\ VGG-19~\cite{simonyan2014very} and ResNet-50~\cite{He2015} were taken as the subject.

\begin{figure}[htbp]
\begin{center}
\subfloat[Original ]{
\includegraphics[width=.15\textwidth]{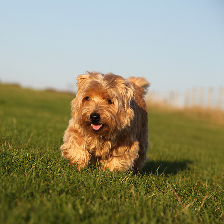}}
\hspace{2em}
\subfloat[Downsampled ]{
\includegraphics[width=.15\textwidth]{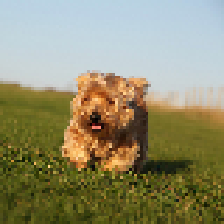}} 
\caption{Mutation of the ImageNet test data}
\label{fig:downsampling}
\end{center}
\end{figure}

Again, Experiments 13 and 14 used the original test set as the operational data. 
Experiments 15 and 16 used low resolution images as the operational testing data, 
which were obtained by downsampling the images in the original test set, as shown in Figure~\ref{fig:downsampling}.

%\begin{figure}[!ht]
%\centering
%\subfloat{
%\includegraphics[width=.18\textwidth]{figures/orig1.png}}
%\subfloat{
%\includegraphics[width=.18\textwidth]{figures/blur1.png}} \\
%\subfloat{
%\includegraphics[width=.18\textwidth]{figures/orig2.png}}
%\subfloat{
%\includegraphics[width=.18\textwidth]{figures/blur2.png}}
%\caption{Mutation of the ImageNet test data}
%\label{fig:downsampling}
%\end{figure}

\begin{figure*}[ht] 
\centering 
\subfloat[Original dataset. VGG-19]{
\includegraphics[width=.245\textwidth]{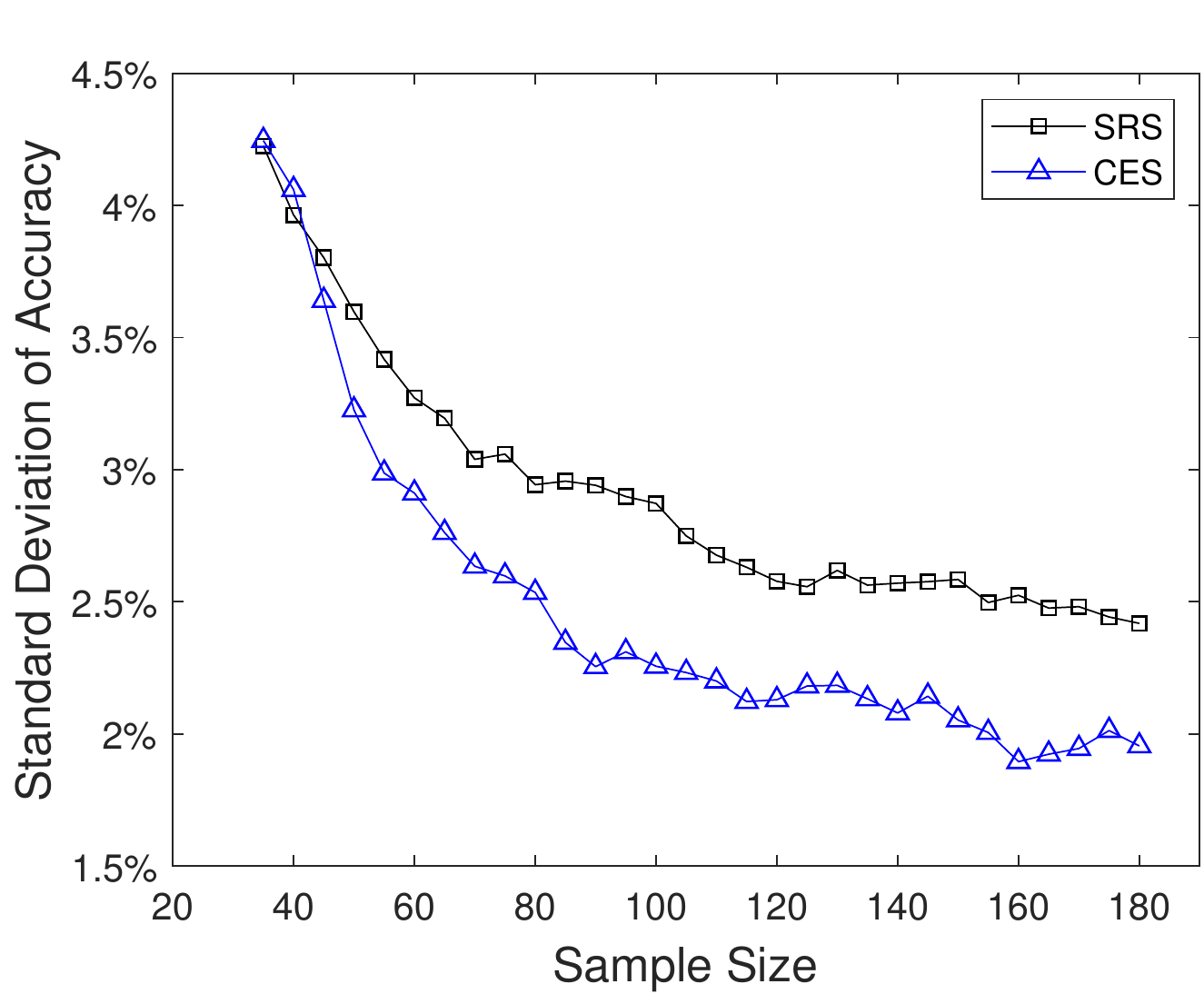}}
\subfloat[Original dataset, ResNet-50]{
\includegraphics[width=.245\textwidth]{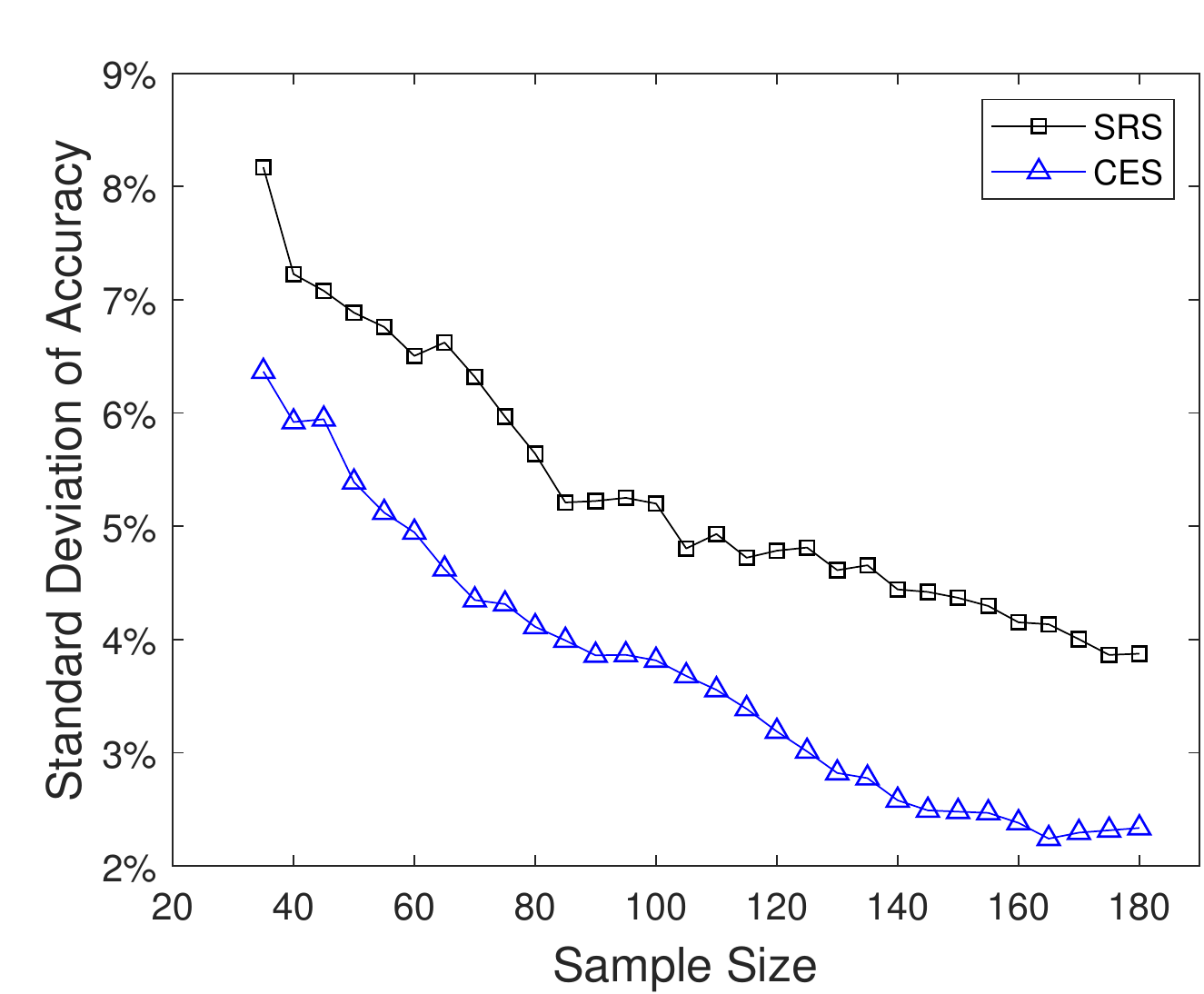}}
\subfloat[Downsampling, VGG-19]{
\includegraphics[width=.245\textwidth]{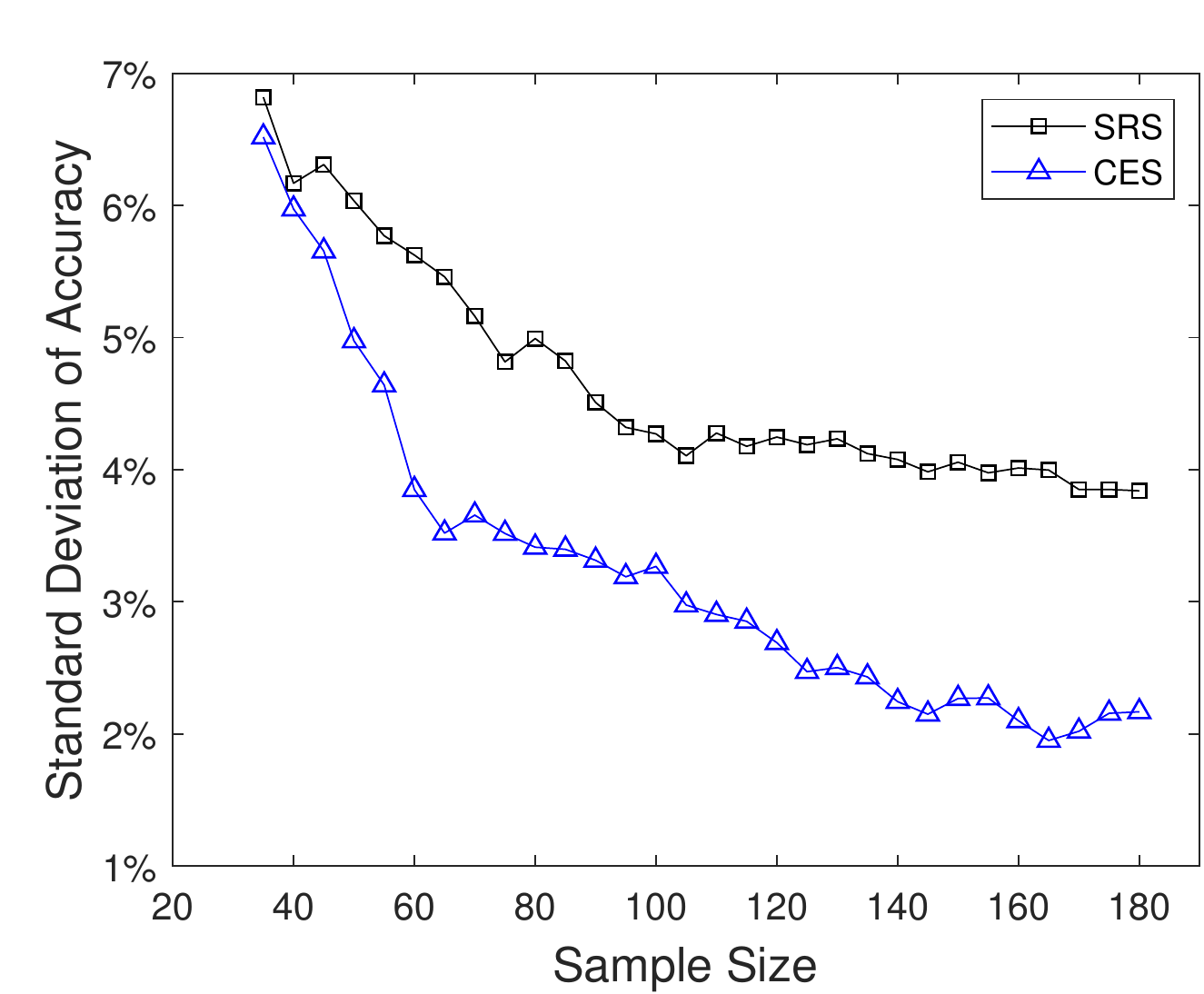}} 
\subfloat[Downsampling, ResNet-50]{
\includegraphics[width=.245\textwidth]{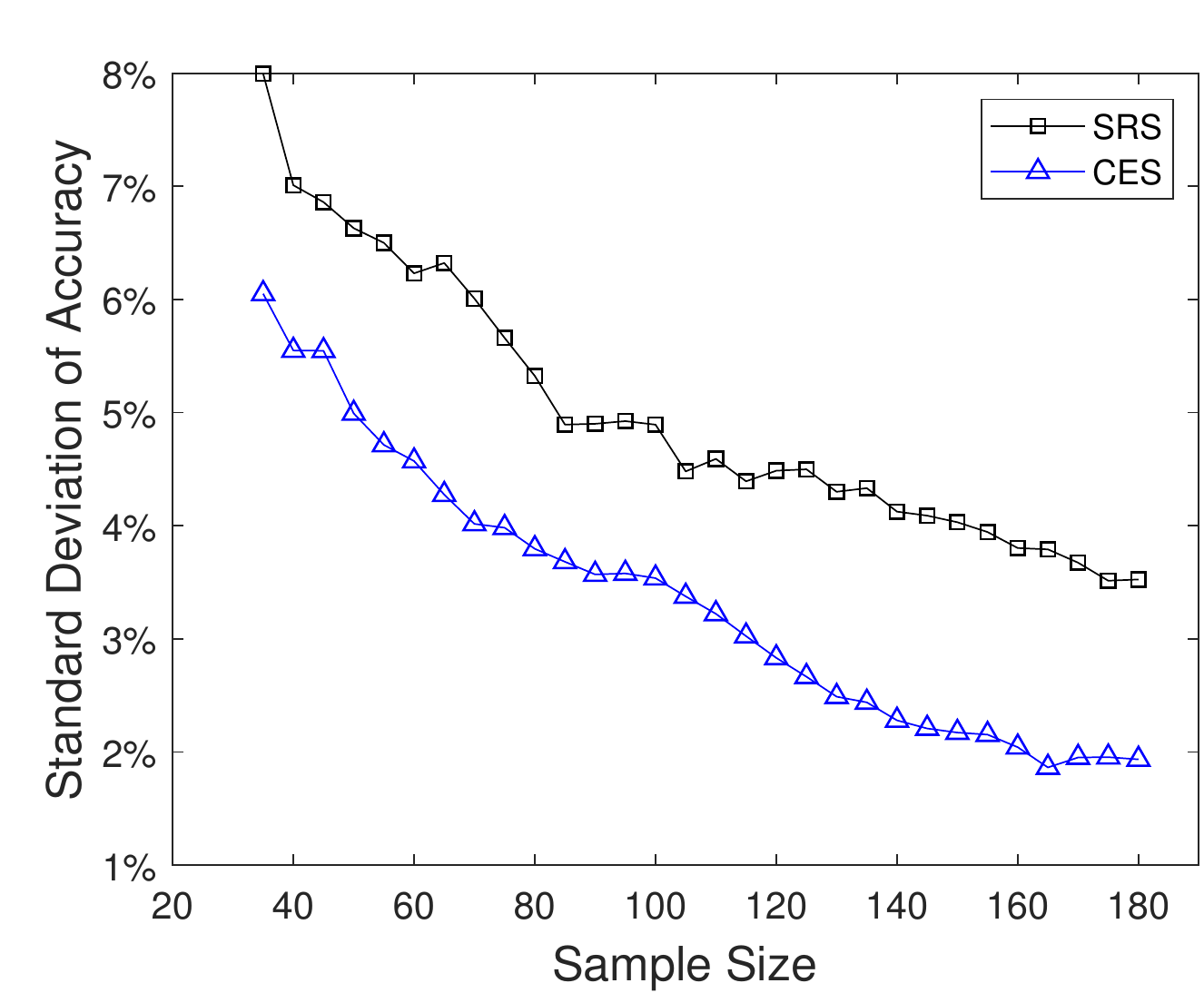}} 
\caption{Results of experiments with ImageNet}     
\label{fig:imagenet}
\end{figure*}

The results of these experiments, as shown in Figure~\ref{fig:imagenet}, are consistent with previous ones. 
It is quite impressive that the CES achieved a standard deviation of a little more than 2\% with only 180 labelled examples,
considering the 1,000 classes of images. 
It indicates that our CES method also performed well for large DNN models, 
with or without the divergence between training data and operational data. 

%MSE are given in Figure , the first two figures are the MSE of estimator based on SRS and CES for original dataset and the last two figures are plotted based on operational dataset.

% the original width=0.38

%Above all, we can conclude the following three points:
%\begin{itemize}
%\item[1.] CES is an effective sampling estimation method. 
%Based on the experiments of the above three datasets, we can find that the CES can save about half of the sample size comparing to the SRS.
%\item[2.] CES has some robustness.
%For changing of datasets or transferring of application environments, the CSS become less efficient and even worse than the SRS. 
%But the CES keep effective enough. 
%\item[3.] 
%\end{itemize}

\subsubsection{Experiments with small operational datasets}
\label{subsubsec:smallopset}
We randomly selected 100 and 300 examples from the MNIST test set and the Driving test set as unlabeled operational test set.
Then SRS and our CES were applied to reduce them to 30 and 50 for labeling, respectively. 
Considering the small sizes of the operational test sets, for better numerical precision in the calculation of Equation~\ref{equ:crossentropy}, we minimizing $D_{KL}(P_T,P_S)$ instead of $D_{KL}(P_S,P_T)=CE(T)-H(P_S)$.  
%compute $D_{KL}(P_T,P_S)=H(P_T,P_S)-H(P_T)$ instead of $H(P_S,P_T)$. 
We also set the number of initial test set $p=5$ in Algorithm~\ref{alg:1} to fit the small sample sizes.

\begin{table}[htb]
\caption{Relative efficiency in case of small test sets}
\centering
\setlength{\tabcolsep}{.42em}
\begin{tabular}{|c|c|c|c|}
\hline
\textbf{No.} & \textbf{Average} & \textbf{Standard Deviation} & \textbf{Maximum} \\
\hline
17 & 0.443  & 0.154 & 0.103   \\
18 &  0.375 & 0.115 & 0.657    \\
19 & 0.594  &  0.265 & 0.946   \\
20 & 0.549  &  0.162 & 0.903   \\
\hline
\end{tabular}
\label{tab:smallopset}
\end{table}

Each of the experiments was repeated for 30 times, 
and the averages, standard deviations, and maximums of relative efficiency are presented in Table~\ref{tab:smallopset}. 
We can see that, our approach still achieved average relative efficiency of 0.375-0.594.
%, but with more fluctuation which was expected, considering the uncertainties of the small operational test set as the ground truth. 
We also found that in almost all cases our approach outperformed SRS, despite of the expected fluctuation in efficiency.  
The result indicates that our approach worked well even in the case that only a very small operational dataset was available.

%% file: relatedwork.tex
% !TEX root = DistCov.tex

\section{Related work}
\label{sec:RelatedWork}
The operational DNN testing addressed by this paper is in line with 
conventional operational testing for software reliability engineering,
but with different subjects, constraints and solutions. 
The work is also complementary to a line of recent research on DNN testing, 
especially those hunting for adversarial examples leveraging structural coverages. 
In the following, we briefly discuss these related work and highlight how our works differs from them.

\subsection{Operational Software Testing}
\label{subsec:OperationalTesting}
The purpose of software testing for human written programs can be 
either \emph{fault detection}, i.e., to find one or more error-inducing inputs for the program under test, 
or \emph{reliability assessment}, i.e., to estimate the chance that the program will err %on a randomly selected input 
in an operation environment. 
Frankl et al.\ ~\cite{Frankl_1998_TES} named these two kinds of testing \emph{debug testing} and \emph{operational testing}, respectively.    
Both of them are integral parts of Software Reliability Engineering~\cite{Musa-1993-IEEESoftware, Lyu-2007-SRE}, 
but they are different in philosophy and technology.   

Operational testing emphasizes maximal improvement of system reliability in operation with limited testing resources, 
so it focuses more on bugs encountered more often in the operation context.   
Debug testing aims at finding as many bugs as possible without explicit consideration of their occurrences in operation. 
The rationales include the belief that software, as logic products, should be correct in all contexts, 
the unavailability of precise operational profile~\cite{Musa-1993-IEEESoftware}, 
and the need for finding rare bugs for systems demanding high reliability. 
Note that, as discussed in Section~\ref{subsec:DNNTesting}, these rationales do not apply to DNNs. 

Technically, operational testing heavily uses statistics and randomization. 
A common theme is to minimize the variance of the estimator for reliability 
through optimal allocations of testing cases, e.g.,~\cite{Maxim-1977-TR,Huang-2005-TR,Lv_2014_TSE}. 
For debug testing, a frequent research topic is the automatic generation of test cases 
with the information provided by the program itself or its models. 
Particularly, structural coverage criteria are used to guide the test case generation~\cite{Anand2013AnOS,Chen_2008_TOSEM,Gay-2015-TSE}. 

Recently, Russo et al.\ showed that further reliability improvement can be achieved by combining the strengths of 
operational testing and structural coverages~\cite{Cotr_2016_TSE,Bertolino_2017_ICSE}. 
In addition, B\"ohme called for a general statistical framework for software testing, and proposed to view software testing as species discovery in order to borrow results from ecological biostatistics~\cite{Bohme-2019-NIER}.

Our work is similar to conventional operation testing in minimizing estimation variance with limited testing resources. 
But we do not have operation profiles except for the unlabeled operational data. 
Instead of using adaptive estimate-allocate-test iterations, 
we leverage the representation information of the DNN under testing to achieve efficient sampling in one step.  
In a sense our conditioning on representation can be viewed as a generalization of the idea behind structural coverages
in debug testing, as discussed earlier.

\subsection{DNN Testing}

Here we consider the testing of well trained DNN models as software artifacts, 
%and do not consider the validation steps included in the training process~\cite{Goodfellow-2016-DL}. 
but not the validation step in training process~\cite{Goodfellow-2016-DL}.

\subsubsection{Structural coverage criteria} 
Recently there is an increasing interest in software testing of machine learning programs~\cite{Houssem-2018-TestMLProg}.
Especially, some authors proposed several structural coverage criteria for DNN testing, borrowing the concept of structural coverage criteria for human written programs~\cite{DeepGaugeASE18,Pei_2017_SOSP,DBLP:journals/corr/abs-1803-04792,ma2018combinatorial,sun2018concolic}. 
The basic idea, is to generate \emph{artificial} examples to 
cover ``corner'' cases of the DNN model that usual test inputs are unable to touch,
in the spirit of debug testing. 
The efficacy of the criteria were illustrated by the \emph{adversarial} examples found in the testing. 
An adversarial example is a slightly perturbed example that fools the DNN model.  

%
%Recently, borrowing the idea of structural coverage, 
%there emerges  for DNN testing.
%Most of them are all applied to cover the corner case of the DNN model, and reveal them by the \emph{adversarial} examples.
%Here we briefly introduce these coverage criteria.
The criteria can be roughly classified into three types: 
\begin{itemize}[leftmargin=*]
\item \emph{Neuron-Activation coverage}: DeepXplore~\cite{Pei_2017_SOSP} defines the \emph{neuron coverage} of a neural network as the percentage of neurons that are activated by the given test set. It uses this coverage, joint with gradient-based optimization, to search for adversarial examples.
%, which are automatically decided by cross-referencing three DNN models.
The weakness of this criterion is that it can be saturated with a small number of test inputs.  
\item \emph{Neuron-Output coverage}: 
DeepGauge~\cite{DeepGaugeASE18} proposes finer-grained criteria to overcome the weakness. It divides the output of each neuron into \textit{k} equal sections, and defines the \emph{\textit{k}-multisection coverage} as the number of chunks that have been covered by the test set. In addition, the authors also considered the touch of outputs out of the $k$ sections with \emph{neuron bound coverage}. % Their work is very similar to ours, and the main difference are:
\item \emph{Neuron-Combination coverages}: The problem of coverage saturation can also be solved with the combination of neuron states. %Different from the previous two coverage criteria which considering the neuron independently, 
DeepCover~\cite{DBLP:journals/corr/abs-1803-04792} and DeepCT~\cite{ma2018combinatorial} use this strategy.  
Inspired by the MC/DC coverage~\cite{hayhurst2001practical}, DeepCover proposes a family of coverage metrics that are very fine-grained.
%combination of neuron-state (i.e. neurons' activation or output) as guides.
%Take the \emph{SS-cover} of DeepCover as example: 
%A pair of neurons $(n_1,n_2)$ in the adjacent layers is said to be SS-Covered by a pair of inputs $(\bm{x}_1,\bm{x}_2)$ when the following conditions are satisfied:
%\begin{itemize}
%\item $n_1$ has a different sign for inputs $(\boldsymbol{x}_1,\boldsymbol{x}_2)$, i.e. the neuron $n_1$ has flipped between $\boldsymbol{x}_1$ and $\boldsymbol{x}_2$.
%\item All other neurons in the layer containing $n_1$ keep the sign, i.e. the other neurons in this layer has not flipped. 
%\item $n_2$ has a different sign for inputs $(\boldsymbol{x}_1,\boldsymbol{x}_2)$.
%\end{itemize}
\end{itemize}

In addition, a recent proposal measures the surprise value of an input as the difference 
in DNN behavior between the input and the training data, 
and use surprise coverage to guide the search for error-inducing inputs~\cite{Kim2019ICSE}.

%Technically, our work has some similarity to DeepGauge in that we also divide a neuron's output into multiple sections. 
%However, we consider only the neurons in the last hidden layer, 
%and leverage the probabilistic distribution of operational data in these sections. 
 
%More fundamentally,
We are skeptical about structural coverage criteria for DNN debug testing~\cite{li2019Bnier}. 
As explained in Section~\ref{subsec:DNNTesting} and Section~\ref{subsec:OperationalTesting}, 
DNN testing needs to be statistical, holistic and operational. 
In addition, the homogeneity-diversity wisdom of coverage-oriented testing is broken by the 
fact that adversarial examples are pervasively distributed over the input space partitioned by these criteria~\cite{li2019Bnier}. 
 
%Different from these approaches, the purpose of our work is to estimate a DNN's accuracy precisely and efficiently, but not to find individual error-inducing inputs.
%generating artificial inputs,   
%sampling from the unlabeled operational data set, in order to reduce the cost of labeling. 
Despite of the difference in purpose, 
one may wonder whether these coverage measures would help in accuracy estimation. 
We believe these scalar measures are not informative enough for improving sampling efficiency. 
To verify this, we experimented with the surprise value in a similar way as CSS, 
and it turned out to be ineffective. The result can be found at our code website. 

\subsubsection{DNN testing for other purposes}

DNN testing is also conducted for purposes other than searching for adversarial examples~\cite{goodfellow2014explaining}. For example,
\begin{itemize}
%In addition to these critera which measure the coverage of corner case.
\item TensorFuzz~\cite{odena2018tensorfuzz} develops a coverage-guided fuzzing (CGF) method to quickly find numerical errors in neural networks.%, and expose the disagreements between neural networks and their quantized versions.
%In addition to coverage-guided DNN testing, many specific testing methods for DNN also have been proposed.
\item DeepMutation~\cite{ma2018deepmutation} builds a DNN mutation testing framework. % to measure the quality of test data.
It simulates potential defects of DNNs through training data mutations %such as data repetition, mislabeling, perturbation, shuffle and deletion, 
and program mutation.%, such as layer removal, layer addition, and activation function removal. 
%Data mutation operators include .
%Program mutation operators consist of .
\item MODE~\cite{ma2018mode} conducts model state differential analysis to determine whether a model is over-fitting or under-fitting. It then performs training input selection that is similar to program input selection in regression testing.
\item DeepRoad~\cite{zhang2018deeproad} uses Generative Adversarial Networks (GANs) to automatically generate 
self-driving images in different weather conditions for DNN testing.
%In particular, they generate driving scenes with various (including those with rather extreme conditions) through the .
\end{itemize}
It is part of our future work to investigate whether our technique can be used for these purposes.

%% file: conclusions.tex
\section{Conclusions}
\label{sec:Conclusions}

A crucial premise for a trained DNN model to work well in a specific operation context is that 
the distribution it learned from the training data is consistent with the operation context. 
Operational testing must be carried out to validate this premise before adopting the model. 
Although deep learning is generally considered as an approach relying on ``big'' data, 
the operational testing of DNNs is often constrained by a limited budget for labeling operational examples, 
and thus it must take a ``small'' data approach that is statistically efficient. 

In this paper we exploit the representation learned by the DNN model to boost the efficiency of operational DNN testing. 
It is interesting to see that although we cannot trust any result produced by the model beforehand 
(recall the unreliability of confidence in Section~\ref{subsec:condconf}), 
%which is demonstrated by the unreliability of confidence value, 
we can still make use of its ``reasoning'', despite of its opaqueness. 
The empirical evaluation confirmed the general efficacy of our approach based on conditioning on representation, 
which reduced the number of labeled operational examples by about a half.   

Another interesting observation is that the homogeneity-diversity wisdom of structural coverage guided testing 
can be generalized to conditioning for variance reduction in reliability estimation. 
The importance of this generalization lies in its potential application to the testing of 
hybrid systems consisting of both DNNs and human written programs. 

There are many possible optimizations left for future work, 
such as further variance reduction through adaptive importance sampling and reducing the volume of unlabeled operational data. 
However, the most important thing to us is how to reformulate the concept of ``bug'' and ``debug", in a statistical, holistic, and operational way that is required for DNNs. 

%% file: DistCov.bbl

\begin{thebibliography}{47}

%%% the Web address.

\ifx \showCODEN    \undefined \def \showCODEN     #1{\unskip}     \fi
\ifx \showDOI      \undefined \def \showDOI       #1{#1}\fi
\ifx \showISBNx    \undefined \def \showISBNx     #1{\unskip}     \fi
\ifx \showISBNxiii \undefined \def \showISBNxiii  #1{\unskip}     \fi
\ifx \showISSN     \undefined \def \showISSN      #1{\unskip}     \fi
\ifx \showLCCN     \undefined \def \showLCCN      #1{\unskip}     \fi
\ifx \shownote     \undefined \def \shownote      #1{#1}          \fi
\ifx \showarticletitle \undefined \def \showarticletitle #1{#1}   \fi
\ifx \showURL      \undefined \def \showURL       {\relax}        \fi
% invisible to TeX
\providecommand\bibfield[2]{#2}
\providecommand\bibinfo[2]{#2}
\providecommand\natexlab[1]{#1}
\providecommand\showeprint[2][]{arXiv:#2}

\bibitem[\protect\citeauthoryear{Abid, Ghorbani, and Zou}{Abid
  et~al\mbox{.}}{2019}]%
        {Zou-2019-AAAI-Fragile}
\bibfield{author}{\bibinfo{person}{Abubakar Abid}, \bibinfo{person}{Amirata
  Ghorbani}, {and} \bibinfo{person}{James Zou}.}
  \bibinfo{year}{2019}\natexlab{}.
\newblock \showarticletitle{Interpretation of Neural Networks is Fragile}. In
  \bibinfo{booktitle}{\emph{Proceedings of the 33rd AAAI Conference on
  Artificial Intelligence (AAAI '19)}}. \bibinfo{address}{Honolulu, Huawaii,
  USA}.
\newblock


\bibitem[\protect\citeauthoryear{Anand, Burke, Chen, Clark, Cohen, Grieskamp,
  Harman, Harrold, and McMinn}{Anand et~al\mbox{.}}{2013}]%
        {Anand2013AnOS}
\bibfield{author}{\bibinfo{person}{Saswat Anand}, \bibinfo{person}{Edmund~K.
  Burke}, \bibinfo{person}{Tsong~Yueh Chen}, \bibinfo{person}{John~A. Clark},
  \bibinfo{person}{Myra~B. Cohen}, \bibinfo{person}{Wolfgang Grieskamp},
  \bibinfo{person}{Mark Harman}, \bibinfo{person}{Mary~Jean Harrold}, {and}
  \bibinfo{person}{Phil McMinn}.} \bibinfo{year}{2013}\natexlab{}.
\newblock \showarticletitle{An orchestrated survey of methodologies for
  automated software test case generation}.
\newblock \bibinfo{journal}{\emph{Journal of Systems and Software}}
  \bibinfo{volume}{86} (\bibinfo{year}{2013}), \bibinfo{pages}{1978--2001}.
\newblock


\bibitem[\protect\citeauthoryear{Beleites, Neugebauer, Bocklitz, Krafft, and
  Popp}{Beleites et~al\mbox{.}}{2013}]%
        {Beleites-2013-ACA}
\bibfield{author}{\bibinfo{person}{Claudia Beleites}, \bibinfo{person}{Ute
  Neugebauer}, \bibinfo{person}{Thomas Bocklitz}, \bibinfo{person}{Christoph
  Krafft}, {and} \bibinfo{person}{J{\"u}rgen Popp}.}
  \bibinfo{year}{2013}\natexlab{}.
\newblock \showarticletitle{Sample size planning for classification models}.
\newblock \bibinfo{journal}{\emph{Analytica Chimica Acta}}
  \bibinfo{volume}{760} (\bibinfo{year}{2013}), \bibinfo{pages}{25 -- 33}.
\newblock
\showISSN{0003-2670}
\urldef\tempurl%
\url{https://doi.org/10.1016/j.aca.2012.11.007}
\showDOI{\tempurl}


\bibitem[\protect\citeauthoryear{{Ben Braiek} and {Khomh}}{{Ben Braiek} and
  {Khomh}}{2018}]%
        {Houssem-2018-TestMLProg}
\bibfield{author}{\bibinfo{person}{Houssem {Ben Braiek}} {and}
  \bibinfo{person}{Foutse {Khomh}}.} \bibinfo{year}{2018}\natexlab{}.
\newblock \showarticletitle{{On Testing Machine Learning Programs}}.
\newblock \bibinfo{journal}{\emph{arXiv e-prints}}, Article
  \bibinfo{articleno}{arXiv:1812.02257} (\bibinfo{date}{Dec.}
  \bibinfo{year}{2018}), \bibinfo{numpages}{arXiv:1812.02257}~pages.
\newblock
\showeprint[arxiv]{cs.SE/1812.02257}


\bibitem[\protect\citeauthoryear{Bengio, Mesnil, Dauphin, and Rifai}{Bengio
  et~al\mbox{.}}{2013}]%
        {bengio2013better}
\bibfield{author}{\bibinfo{person}{Yoshua Bengio},
  \bibinfo{person}{Gr{\'e}goire Mesnil}, \bibinfo{person}{Yann Dauphin}, {and}
  \bibinfo{person}{Salah Rifai}.} \bibinfo{year}{2013}\natexlab{}.
\newblock \showarticletitle{Better mixing via deep representations}. In
  \bibinfo{booktitle}{\emph{International conference on machine learning}}.
  \bibinfo{pages}{552--560}.
\newblock


\bibitem[\protect\citeauthoryear{Bertolino, Miranda, Pietrantuono, and
  Russo}{Bertolino et~al\mbox{.}}{2017}]%
        {Bertolino_2017_ICSE}
\bibfield{author}{\bibinfo{person}{Antonia Bertolino}, \bibinfo{person}{Breno
  Miranda}, \bibinfo{person}{Roberto Pietrantuono}, {and}
  \bibinfo{person}{Stefano Russo}.} \bibinfo{year}{2017}\natexlab{}.
\newblock \showarticletitle{Adaptive Coverage and Operational Profile-Based
  Testing for Reliability Improvement}.
\newblock \bibinfo{journal}{\emph{2017 IEEE/ACM 39th International Conference
  on Software Engineering (ICSE)}} (\bibinfo{date}{May} \bibinfo{year}{2017}).
\newblock
\showISBNx{9781538638682}
\urldef\tempurl%
\url{https://doi.org/10.1109/icse.2017.56}
\showDOI{\tempurl}


\bibitem[\protect\citeauthoryear{B{\"o}hme}{B{\"o}hme}{2019}]%
        {Bohme-2019-NIER}
\bibfield{author}{\bibinfo{person}{Marcel B{\"o}hme}.}
  \bibinfo{year}{2019}\natexlab{}.
\newblock \showarticletitle{Assurance in Software Testing: A Roadmap}. In
  \bibinfo{booktitle}{\emph{Proceedings of the 41st International Conference on
  Software Engineering: New Ideas and Emerging Results Track}}.
\newblock


\bibitem[\protect\citeauthoryear{Bojarski, Testa, Dworakowski, Firner, Flepp,
  Goyal, Jackel, Monfort, Muller, Zhang, Zhang, Zhao, and Zieba}{Bojarski
  et~al\mbox{.}}{2016}]%
        {bojarski2016ArXiv}
\bibfield{author}{\bibinfo{person}{Mariusz Bojarski},
  \bibinfo{person}{Davide~Del Testa}, \bibinfo{person}{Daniel Dworakowski},
  \bibinfo{person}{Bernhard Firner}, \bibinfo{person}{Beat Flepp},
  \bibinfo{person}{Prasoon Goyal}, \bibinfo{person}{Lawrence~D. Jackel},
  \bibinfo{person}{Mathew Monfort}, \bibinfo{person}{Urs Muller},
  \bibinfo{person}{Jiakai Zhang}, \bibinfo{person}{Xin Zhang},
  \bibinfo{person}{Jake Zhao}, {and} \bibinfo{person}{Karol Zieba}.}
  \bibinfo{year}{2016}\natexlab{}.
\newblock \showarticletitle{End to End Learning for Self-Driving Cars}.
\newblock \bibinfo{journal}{\emph{CoRR}}  \bibinfo{volume}{abs/1604.07316}
  (\bibinfo{year}{2016}).
\newblock
\showeprint[arxiv]{1604.07316}
\urldef\tempurl%
\url{http://arxiv.org/abs/1604.07316}
\showURL{%
\tempurl}


\bibitem[\protect\citeauthoryear{Chen, Tse, and Yu}{Chen et~al\mbox{.}}{2001}]%
        {Chen_2001_JSS}
\bibfield{author}{\bibinfo{person}{T.Y. Chen}, \bibinfo{person}{T.H. Tse},
  {and} \bibinfo{person}{Y.T. Yu}.} \bibinfo{year}{2001}\natexlab{}.
\newblock \showarticletitle{Proportional sampling strategy: a compendium and
  some insights}.
\newblock \bibinfo{journal}{\emph{Journal of Systems and Software}}
  \bibinfo{volume}{58}, \bibinfo{number}{1} (\bibinfo{date}{Aug}
  \bibinfo{year}{2001}), \bibinfo{pages}{65--81}.
\newblock
\showISSN{0164-1212}
\urldef\tempurl%
\url{https://doi.org/10.1016/s0164-1212(01)00028-0}
\showDOI{\tempurl}


\bibitem[\protect\citeauthoryear{Chen and Merkel}{Chen and Merkel}{2008}]%
        {Chen_2008_TOSEM}
\bibfield{author}{\bibinfo{person}{Tsong~Yueh Chen} {and}
  \bibinfo{person}{Robert Merkel}.} \bibinfo{year}{2008}\natexlab{}.
\newblock \showarticletitle{An upper bound on software testing effectiveness}.
\newblock \bibinfo{journal}{\emph{ACM Transactions on Software Engineering and
  Methodology}} \bibinfo{volume}{17}, \bibinfo{number}{3} (\bibinfo{date}{Jun}
  \bibinfo{year}{2008}), \bibinfo{pages}{1--27}.
\newblock
\showISSN{1049-331X}
\urldef\tempurl%
\url{https://doi.org/10.1145/1363102.1363107}
\showDOI{\tempurl}


\bibitem[\protect\citeauthoryear{Cotroneo, Pietrantuono, and Russo}{Cotroneo
  et~al\mbox{.}}{2016}]%
        {Cotr_2016_TSE}
\bibfield{author}{\bibinfo{person}{Domenico Cotroneo}, \bibinfo{person}{Roberto
  Pietrantuono}, {and} \bibinfo{person}{Stefano Russo}.}
  \bibinfo{year}{2016}\natexlab{}.
\newblock \showarticletitle{RELAI Testing: A Technique to Assess and Improve
  Software Reliability}.
\newblock \bibinfo{journal}{\emph{IEEE Transactions on Software Engineering}}
  \bibinfo{volume}{42}, \bibinfo{number}{5} (\bibinfo{date}{May}
  \bibinfo{year}{2016}), \bibinfo{pages}{452--475}.
\newblock
\showISSN{0098-5589}
\urldef\tempurl%
\url{https://doi.org/10.1109/TSE.2015.2491931}
\showDOI{\tempurl}


\bibitem[\protect\citeauthoryear{Deng, Dong, Socher, Li, Li, and Fei-Fei}{Deng
  et~al\mbox{.}}{2009}]%
        {imagenet_cvpr09}
\bibfield{author}{\bibinfo{person}{J. Deng}, \bibinfo{person}{W. Dong},
  \bibinfo{person}{R. Socher}, \bibinfo{person}{L.-J. Li}, \bibinfo{person}{K.
  Li}, {and} \bibinfo{person}{L. Fei-Fei}.} \bibinfo{year}{2009}\natexlab{}.
\newblock \showarticletitle{{ImageNet: A Large-Scale Hierarchical Image
  Database}}. In \bibinfo{booktitle}{\emph{CVPR09}}.
\newblock


\bibitem[\protect\citeauthoryear{Frankl, Hamlet, Littlewood, and
  Strigini}{Frankl et~al\mbox{.}}{1998}]%
        {Frankl_1998_TES}
\bibfield{author}{\bibinfo{person}{Phyllis~G. Frankl},
  \bibinfo{person}{Richard~G. Hamlet}, \bibinfo{person}{Bev Littlewood}, {and}
  \bibinfo{person}{Lorenzo Strigini}.} \bibinfo{year}{1998}\natexlab{}.
\newblock \showarticletitle{Evaluating Testing Methods by Delivered
  Reliability}.
\newblock \bibinfo{journal}{\emph{IEEE Trans. Softw. Eng.}}
  \bibinfo{volume}{24}, \bibinfo{number}{8} (\bibinfo{date}{Aug.}
  \bibinfo{year}{1998}), \bibinfo{pages}{586--601}.
\newblock
\showISSN{0098-5589}
\urldef\tempurl%
\url{https://doi.org/10.1109/32.707695}
\showDOI{\tempurl}


\bibitem[\protect\citeauthoryear{Gay, Staats, Whalen, and Heimdahl}{Gay
  et~al\mbox{.}}{2015}]%
        {Gay-2015-TSE}
\bibfield{author}{\bibinfo{person}{Gregory Gay}, \bibinfo{person}{Matt Staats},
  \bibinfo{person}{Michael Whalen}, {and} \bibinfo{person}{Mats P.~E.
  Heimdahl}.} \bibinfo{year}{2015}\natexlab{}.
\newblock \showarticletitle{The Risks of Coverage-Directed Test Case
  Generation}.
\newblock \bibinfo{journal}{\emph{IEEE Transactions on Software Engineering}}
  \bibinfo{volume}{41}, \bibinfo{number}{8} (\bibinfo{date}{August}
  \bibinfo{year}{2015}), \bibinfo{pages}{803--819}.
\newblock


\bibitem[\protect\citeauthoryear{Goodfellow, Bengio, and Courville}{Goodfellow
  et~al\mbox{.}}{2016}]%
        {Goodfellow-2016-DL}
\bibfield{author}{\bibinfo{person}{Ian Goodfellow}, \bibinfo{person}{Yoshua
  Bengio}, {and} \bibinfo{person}{Aaron Courville}.}
  \bibinfo{year}{2016}\natexlab{}.
\newblock \bibinfo{booktitle}{\emph{Deep Learning}}.
\newblock \bibinfo{publisher}{The MIT Press}.
\newblock
\showISBNx{0262035618, 9780262035613}


\bibitem[\protect\citeauthoryear{Goodfellow, Shlens, and Szegedy}{Goodfellow
  et~al\mbox{.}}{2014}]%
        {goodfellow2014explaining}
\bibfield{author}{\bibinfo{person}{Ian~J Goodfellow}, \bibinfo{person}{Jonathon
  Shlens}, {and} \bibinfo{person}{Christian Szegedy}.}
  \bibinfo{year}{2014}\natexlab{}.
\newblock \showarticletitle{Explaining and harnessing adversarial examples}.
\newblock \bibinfo{journal}{\emph{arXiv preprint arXiv:1412.6572}}
  (\bibinfo{year}{2014}).
\newblock


\bibitem[\protect\citeauthoryear{Hayhurst, Veerhusen, Chilenski, and
  Rierson}{Hayhurst et~al\mbox{.}}{2001}]%
        {hayhurst2001practical}
\bibfield{author}{\bibinfo{person}{Kelly~J Hayhurst}, \bibinfo{person}{Dan~S
  Veerhusen}, \bibinfo{person}{John~J Chilenski}, {and}
  \bibinfo{person}{Leanna~K Rierson}.} \bibinfo{year}{2001}\natexlab{}.
\newblock \showarticletitle{A practical tutorial on modified condition/decision
  coverage}.
\newblock  (\bibinfo{year}{2001}).
\newblock


\bibitem[\protect\citeauthoryear{He, Zhang, Ren, and Sun}{He
  et~al\mbox{.}}{2015}]%
        {He2015}
\bibfield{author}{\bibinfo{person}{Kaiming He}, \bibinfo{person}{Xiangyu
  Zhang}, \bibinfo{person}{Shaoqing Ren}, {and} \bibinfo{person}{Jian Sun}.}
  \bibinfo{year}{2015}\natexlab{}.
\newblock \showarticletitle{Deep Residual Learning for Image Recognition}.
\newblock \bibinfo{journal}{\emph{arXiv preprint arXiv:1512.03385}}
  (\bibinfo{year}{2015}).
\newblock


\bibitem[\protect\citeauthoryear{Ho and Yeung}{Ho and Yeung}{2010}]%
        {ho2010information}
\bibfield{author}{\bibinfo{person}{Siu-Wai Ho} {and} \bibinfo{person}{Raymond~W
  Yeung}.} \bibinfo{year}{2010}\natexlab{}.
\newblock \showarticletitle{On information divergence measures and a unified
  typicality}.
\newblock \bibinfo{journal}{\emph{IEEE Transactions on Information Theory}}
  \bibinfo{volume}{56}, \bibinfo{number}{12} (\bibinfo{year}{2010}),
  \bibinfo{pages}{5893--5905}.
\newblock


\bibitem[\protect\citeauthoryear{Huang and Lyu}{Huang and Lyu}{2005}]%
        {Huang-2005-TR}
\bibfield{author}{\bibinfo{person}{Chin-Yu Huang} {and}
  \bibinfo{person}{Michael~R. Lyu}.} \bibinfo{year}{2005}\natexlab{}.
\newblock \showarticletitle{Optimal Testing Resource Allocation, and
  Sensitivity Analysis in Software Development}.
\newblock \bibinfo{journal}{\emph{IEEE Transactions on Reliability}}
  \bibinfo{volume}{54}, \bibinfo{number}{4} (\bibinfo{date}{December}
  \bibinfo{year}{2005}), \bibinfo{pages}{592--603}.
\newblock


\bibitem[\protect\citeauthoryear{{Huang}, {Li}, {Yu}, {Deng}, and
  {Gong}}{{Huang} et~al\mbox{.}}{2013}]%
        {Huang-2013-ICASSP}
\bibfield{author}{\bibinfo{person}{Jui-Ting {Huang}}, \bibinfo{person}{Jinyu
  {Li}}, \bibinfo{person}{Dong {Yu}}, \bibinfo{person}{Li {Deng}}, {and}
  \bibinfo{person}{Yifan {Gong}}.} \bibinfo{year}{2013}\natexlab{}.
\newblock \showarticletitle{Cross-language knowledge transfer using
  multilingual deep neural network with shared hidden layers}. In
  \bibinfo{booktitle}{\emph{2013 IEEE International Conference on Acoustics,
  Speech and Signal Processing}}. \bibinfo{pages}{7304--7308}.
\newblock
\showISSN{1520-6149}
\urldef\tempurl%
\url{https://doi.org/10.1109/ICASSP.2013.6639081}
\showDOI{\tempurl}


\bibitem[\protect\citeauthoryear{Kim, Feldt, and Yoo}{Kim
  et~al\mbox{.}}{2019}]%
        {Kim2019ICSE}
\bibfield{author}{\bibinfo{person}{Jinhan Kim}, \bibinfo{person}{Robert Feldt},
  {and} \bibinfo{person}{Shin Yoo}.} \bibinfo{year}{2019}\natexlab{}.
\newblock \showarticletitle{Guiding Deep Learning System Testing Using Surprise
  Adequacy}. In \bibinfo{booktitle}{\emph{Proceedings of the 41st International
  Conference on Software Engineering}} \emph{(\bibinfo{series}{ICSE '19})}.
  \bibinfo{publisher}{IEEE Press}, \bibinfo{address}{Piscataway, NJ, USA},
  \bibinfo{pages}{1039--1049}.
\newblock
\urldef\tempurl%
\url{https://doi.org/10.1109/ICSE.2019.00108}
\showDOI{\tempurl}


\bibitem[\protect\citeauthoryear{Koller, Friedman, and Bach}{Koller
  et~al\mbox{.}}{2009}]%
        {koller2009probabilistic}
\bibfield{author}{\bibinfo{person}{Daphne Koller}, \bibinfo{person}{Nir
  Friedman}, {and} \bibinfo{person}{Francis Bach}.}
  \bibinfo{year}{2009}\natexlab{}.
\newblock \bibinfo{booktitle}{\emph{Probabilistic graphical models: principles
  and techniques}}.
\newblock \bibinfo{publisher}{MIT press}.
\newblock


\bibitem[\protect\citeauthoryear{LeCun, Bengio, and Hinton}{LeCun
  et~al\mbox{.}}{2015}]%
        {lecun2015Nature-DeepLearning}
\bibfield{author}{\bibinfo{person}{Yann LeCun}, \bibinfo{person}{Yoshua
  Bengio}, {and} \bibinfo{person}{Geoffrey Hinton}.}
  \bibinfo{year}{2015}\natexlab{}.
\newblock \showarticletitle{Deep learning}.
\newblock \bibinfo{journal}{\emph{Nature}} \bibinfo{volume}{521},
  \bibinfo{number}{7553} (\bibinfo{year}{2015}), \bibinfo{pages}{436--444}.
\newblock


\bibitem[\protect\citeauthoryear{LeCun, Bottou, Bengio, Haffner,
  et~al\mbox{.}}{LeCun et~al\mbox{.}}{1998}]%
        {lecun1998gradient}
\bibfield{author}{\bibinfo{person}{Yann LeCun}, \bibinfo{person}{L{\'e}on
  Bottou}, \bibinfo{person}{Yoshua Bengio}, \bibinfo{person}{Patrick Haffner},
  {et~al\mbox{.}}} \bibinfo{year}{1998}\natexlab{}.
\newblock \showarticletitle{Gradient-based learning applied to document
  recognition}.
\newblock \bibinfo{journal}{\emph{Proc. IEEE}} \bibinfo{volume}{86},
  \bibinfo{number}{11} (\bibinfo{year}{1998}), \bibinfo{pages}{2278--2324}.
\newblock


\bibitem[\protect\citeauthoryear{Li, Ma, Xu, and Cao}{Li et~al\mbox{.}}{2019}]%
        {li2019Bnier}
\bibfield{author}{\bibinfo{person}{Zenan Li}, \bibinfo{person}{Xiaoxing Ma},
  \bibinfo{person}{Chang Xu}, {and} \bibinfo{person}{Chun Cao}.}
  \bibinfo{year}{2019}\natexlab{}.
\newblock \showarticletitle{Structural Coverage Criteria for Neural Networks
  Could Be Misleading}. In \bibinfo{booktitle}{\emph{Proceedings of the 41st
  International Conference on Software Engineering: New Ideas and Emerging
  Results}} \emph{(\bibinfo{series}{ICSE-NIER '19})}. \bibinfo{publisher}{IEEE
  Press}, \bibinfo{address}{Piscataway, NJ, USA}, \bibinfo{pages}{89--92}.
\newblock
\urldef\tempurl%
\url{https://doi.org/10.1109/ICSE-NIER.2019.00031}
\showDOI{\tempurl}


\bibitem[\protect\citeauthoryear{{Lipton}}{{Lipton}}{2016}]%
        {Lipton-2016-arXiv-Mythos}
\bibfield{author}{\bibinfo{person}{Zachary~C. {Lipton}}.}
  \bibinfo{year}{2016}\natexlab{}.
\newblock \showarticletitle{{The Mythos of Model Interpretability}}.
\newblock \bibinfo{journal}{\emph{arXiv e-prints}}, Article
  \bibinfo{articleno}{arXiv:1606.03490} (\bibinfo{date}{June}
  \bibinfo{year}{2016}), \bibinfo{numpages}{arXiv:1606.03490}~pages.
\newblock
\showeprint[arxiv]{cs.LG/1606.03490}


\bibitem[\protect\citeauthoryear{Lv, Yin, and Cai}{Lv et~al\mbox{.}}{2014}]%
        {Lv_2014_TSE}
\bibfield{author}{\bibinfo{person}{Junpeng Lv}, \bibinfo{person}{Bei-Bei Yin},
  {and} \bibinfo{person}{Kai-Yuan Cai}.} \bibinfo{year}{2014}\natexlab{}.
\newblock \showarticletitle{On the Asymptotic Behavior of Adaptive Testing
  Strategy for Software Reliability Assessment}.
\newblock \bibinfo{journal}{\emph{IEEE Transactions on Software Engineering}}
  \bibinfo{volume}{40}, \bibinfo{number}{4} (\bibinfo{date}{Apr}
  \bibinfo{year}{2014}), \bibinfo{pages}{396--412}.
\newblock
\showISSN{1939-3520}
\urldef\tempurl%
\url{https://doi.org/10.1109/tse.2014.2310194}
\showDOI{\tempurl}


\bibitem[\protect\citeauthoryear{{Lyu}}{{Lyu}}{2007}]%
        {Lyu-2007-SRE}
\bibfield{author}{\bibinfo{person}{Michael~R. {Lyu}}.}
  \bibinfo{year}{2007}\natexlab{}.
\newblock \showarticletitle{Software Reliability Engineering: A Roadmap}. In
  \bibinfo{booktitle}{\emph{Future of Software Engineering (FOSE '07)}}.
  \bibinfo{pages}{153--170}.
\newblock
\urldef\tempurl%
\url{https://doi.org/10.1109/FOSE.2007.24}
\showDOI{\tempurl}


\bibitem[\protect\citeauthoryear{Ma, Juefei-Xu, Zhang, Sun, Xue, Li, Chen, Su,
  Li, Liu, Zhao, and Wang}{Ma et~al\mbox{.}}{2018a}]%
        {DeepGaugeASE18}
\bibfield{author}{\bibinfo{person}{Lei Ma}, \bibinfo{person}{Felix Juefei-Xu},
  \bibinfo{person}{Fuyuan Zhang}, \bibinfo{person}{Jiyuan Sun},
  \bibinfo{person}{Minhui Xue}, \bibinfo{person}{Bo Li},
  \bibinfo{person}{Chunyang Chen}, \bibinfo{person}{Ting Su},
  \bibinfo{person}{Li Li}, \bibinfo{person}{Yang Liu}, \bibinfo{person}{Jianjun
  Zhao}, {and} \bibinfo{person}{Yadong Wang}.}
  \bibinfo{year}{2018}\natexlab{a}.
\newblock \showarticletitle{DeepGauge: Multi-granularity Testing Criteria for
  Deep Learning Systems}. In \bibinfo{booktitle}{\emph{Proceedings of the 33rd
  ACM/IEEE International Conference on Automated Software Engineering}}
  \emph{(\bibinfo{series}{ASE 2018})}. \bibinfo{publisher}{ACM},
  \bibinfo{address}{New York, NY, USA}, \bibinfo{pages}{120--131}.
\newblock
\showISBNx{978-1-4503-5937-5}
\urldef\tempurl%
\url{https://doi.org/10.1145/3238147.3238202}
\showDOI{\tempurl}


\bibitem[\protect\citeauthoryear{Ma, Zhang, Sun, Xue, Li, Juefei-Xu, Xie, Li,
  Liu, Zhao, et~al\mbox{.}}{Ma et~al\mbox{.}}{2018c}]%
        {ma2018deepmutation}
\bibfield{author}{\bibinfo{person}{Lei Ma}, \bibinfo{person}{Fuyuan Zhang},
  \bibinfo{person}{Jiyuan Sun}, \bibinfo{person}{Minhui Xue},
  \bibinfo{person}{Bo Li}, \bibinfo{person}{Felix Juefei-Xu},
  \bibinfo{person}{Chao Xie}, \bibinfo{person}{Li Li}, \bibinfo{person}{Yang
  Liu}, \bibinfo{person}{Jianjun Zhao}, {et~al\mbox{.}}}
  \bibinfo{year}{2018}\natexlab{c}.
\newblock \showarticletitle{Deepmutation: Mutation testing of deep learning
  systems}. In \bibinfo{booktitle}{\emph{2018 IEEE 29th International Symposium
  on Software Reliability Engineering (ISSRE)}}. IEEE,
  \bibinfo{pages}{100--111}.
\newblock


\bibitem[\protect\citeauthoryear{Ma, Zhang, Xue, Li, Liu, Zhao, and Wang}{Ma
  et~al\mbox{.}}{2018d}]%
        {ma2018combinatorial}
\bibfield{author}{\bibinfo{person}{Lei Ma}, \bibinfo{person}{Fuyuan Zhang},
  \bibinfo{person}{Minhui Xue}, \bibinfo{person}{Bo Li}, \bibinfo{person}{Yang
  Liu}, \bibinfo{person}{Jianjun Zhao}, {and} \bibinfo{person}{Yadong Wang}.}
  \bibinfo{year}{2018}\natexlab{d}.
\newblock \showarticletitle{Combinatorial testing for deep learning systems}.
\newblock \bibinfo{journal}{\emph{arXiv preprint arXiv:1806.07723}}
  (\bibinfo{year}{2018}).
\newblock


\bibitem[\protect\citeauthoryear{Ma, Liu, Lee, Zhang, and Grama}{Ma
  et~al\mbox{.}}{2018b}]%
        {ma2018mode}
\bibfield{author}{\bibinfo{person}{Shiqing Ma}, \bibinfo{person}{Yingqi Liu},
  \bibinfo{person}{Wen-Chuan Lee}, \bibinfo{person}{Xiangyu Zhang}, {and}
  \bibinfo{person}{Ananth Grama}.} \bibinfo{year}{2018}\natexlab{b}.
\newblock \showarticletitle{MODE: automated neural network model debugging via
  state differential analysis and input selection}. In
  \bibinfo{booktitle}{\emph{Proceedings of the 2018 26th ACM Joint Meeting on
  European Software Engineering Conference and Symposium on the Foundations of
  Software Engineering}}. ACM, \bibinfo{pages}{175--186}.
\newblock


\bibitem[\protect\citeauthoryear{MacKay and Mac~Kay}{MacKay and
  Mac~Kay}{2003}]%
        {mackay2003information}
\bibfield{author}{\bibinfo{person}{David~JC MacKay} {and}
  \bibinfo{person}{David~JC Mac~Kay}.} \bibinfo{year}{2003}\natexlab{}.
\newblock \bibinfo{booktitle}{\emph{Information theory, inference and learning
  algorithms}}.
\newblock \bibinfo{publisher}{Cambridge university press}.
\newblock


\bibitem[\protect\citeauthoryear{{Maxim} and {Weed}}{{Maxim} and
  {Weed}}{1977}]%
        {Maxim-1977-TR}
\bibfield{author}{\bibinfo{person}{L.~Daniel {Maxim}} {and}
  \bibinfo{person}{Harrison~D. {Weed}}.} \bibinfo{year}{1977}\natexlab{}.
\newblock \showarticletitle{Allocation of Test Effort for Minimum Variance of
  Reliability}.
\newblock \bibinfo{journal}{\emph{IEEE Transactions on Reliability}}
  \bibinfo{volume}{R-26}, \bibinfo{number}{2} (\bibinfo{date}{June}
  \bibinfo{year}{1977}), \bibinfo{pages}{111--115}.
\newblock
\showISSN{0018-9529}
\urldef\tempurl%
\url{https://doi.org/10.1109/TR.1977.5220068}
\showDOI{\tempurl}


\bibitem[\protect\citeauthoryear{Musa}{Musa}{1993}]%
        {Musa-1993-IEEESoftware}
\bibfield{author}{\bibinfo{person}{John~D. Musa}.}
  \bibinfo{year}{1993}\natexlab{}.
\newblock \showarticletitle{Operational profiles in software-reliability
  engineering}.
\newblock \bibinfo{journal}{\emph{IEEE Software}} \bibinfo{volume}{10},
  \bibinfo{number}{2} (\bibinfo{date}{March} \bibinfo{year}{1993}),
  \bibinfo{pages}{14--32}.
\newblock
\showISSN{0740-7459}
\urldef\tempurl%
\url{https://doi.org/10.1109/52.199724}
\showDOI{\tempurl}


\bibitem[\protect\citeauthoryear{Obermeyer and Emanuel}{Obermeyer and
  Emanuel}{2016}]%
        {Obermeyer-2016-NEJM}
\bibfield{author}{\bibinfo{person}{Ziad Obermeyer} {and}
  \bibinfo{person}{Ezekiel~J Emanuel}.} \bibinfo{year}{2016}\natexlab{}.
\newblock \showarticletitle{Predicting the Future - Big Data, Machine Learning,
  and Clinical Medicine}.
\newblock \bibinfo{journal}{\emph{The New England journal of medicine}}
  \bibinfo{volume}{375}, \bibinfo{number}{13} (\bibinfo{date}{09}
  \bibinfo{year}{2016}), \bibinfo{pages}{1216--1219}.
\newblock
\showISBNx{1533-4406; 0028-4793}
\urldef\tempurl%
\url{https://doi.org/10.1056/NEJMp1606181}
\showDOI{\tempurl}


\bibitem[\protect\citeauthoryear{Odena and Goodfellow}{Odena and
  Goodfellow}{2018}]%
        {odena2018tensorfuzz}
\bibfield{author}{\bibinfo{person}{Augustus Odena} {and} \bibinfo{person}{Ian
  Goodfellow}.} \bibinfo{year}{2018}\natexlab{}.
\newblock \showarticletitle{Tensorfuzz: Debugging neural networks with
  coverage-guided fuzzing}.
\newblock \bibinfo{journal}{\emph{arXiv preprint arXiv:1807.10875}}
  (\bibinfo{year}{2018}).
\newblock


\bibitem[\protect\citeauthoryear{Owen}{Owen}{2013}]%
        {Owen-2013-mcbook}
\bibfield{author}{\bibinfo{person}{Art~B. Owen}.}
  \bibinfo{year}{2013}\natexlab{}.
\newblock \bibinfo{booktitle}{\emph{Monte Carlo theory, methods and examples}}.
\newblock \bibinfo{publisher}{https://statweb.stanford.edu/~owen/mc/}.
\newblock


\bibitem[\protect\citeauthoryear{Pei, Cao, Yang, and Jana}{Pei
  et~al\mbox{.}}{2017}]%
        {Pei_2017_SOSP}
\bibfield{author}{\bibinfo{person}{Kexin Pei}, \bibinfo{person}{Yinzhi Cao},
  \bibinfo{person}{Junfeng Yang}, {and} \bibinfo{person}{Suman Jana}.}
  \bibinfo{year}{2017}\natexlab{}.
\newblock \showarticletitle{DeepXplore: Automated Whitebox Testing of Deep
  Learning Systems}. In \bibinfo{booktitle}{\emph{Proceedings of the 26th
  Symposium on Operating Systems Principles}} \emph{(\bibinfo{series}{SOSP
  '17})}. \bibinfo{publisher}{ACM}, \bibinfo{address}{New York, NY, USA},
  \bibinfo{pages}{1--18}.
\newblock
\showISBNx{978-1-4503-5085-3}
\urldef\tempurl%
\url{https://doi.org/10.1145/3132747.3132785}
\showDOI{\tempurl}


\bibitem[\protect\citeauthoryear{R{\'e}v{\'e}sz}{R{\'e}v{\'e}sz}{2005}]%
        {revesz2005random}
\bibfield{author}{\bibinfo{person}{P{\'a}l R{\'e}v{\'e}sz}.}
  \bibinfo{year}{2005}\natexlab{}.
\newblock \bibinfo{booktitle}{\emph{Random walk in random and non-random
  environments}}.
\newblock \bibinfo{publisher}{World Scientific}.
\newblock


\bibitem[\protect\citeauthoryear{Shore and Johnson}{Shore and Johnson}{1980}]%
        {shore1980axiomatic}
\bibfield{author}{\bibinfo{person}{John Shore} {and} \bibinfo{person}{Rodney
  Johnson}.} \bibinfo{year}{1980}\natexlab{}.
\newblock \showarticletitle{Axiomatic derivation of the principle of maximum
  entropy and the principle of minimum cross-entropy}.
\newblock \bibinfo{journal}{\emph{IEEE Transactions on information theory}}
  \bibinfo{volume}{26}, \bibinfo{number}{1} (\bibinfo{year}{1980}),
  \bibinfo{pages}{26--37}.
\newblock


\bibitem[\protect\citeauthoryear{Simonyan and Zisserman}{Simonyan and
  Zisserman}{2014}]%
        {simonyan2014very}
\bibfield{author}{\bibinfo{person}{Karen Simonyan} {and}
  \bibinfo{person}{Andrew Zisserman}.} \bibinfo{year}{2014}\natexlab{}.
\newblock \showarticletitle{Very deep convolutional networks for large-scale
  image recognition}.
\newblock \bibinfo{journal}{\emph{arXiv preprint arXiv:1409.1556}}
  (\bibinfo{year}{2014}).
\newblock


\bibitem[\protect\citeauthoryear{Sun, Huang, and Kroening}{Sun
  et~al\mbox{.}}{2018a}]%
        {DBLP:journals/corr/abs-1803-04792}
\bibfield{author}{\bibinfo{person}{Youcheng Sun}, \bibinfo{person}{Xiaowei
  Huang}, {and} \bibinfo{person}{Daniel Kroening}.}
  \bibinfo{year}{2018}\natexlab{a}.
\newblock \showarticletitle{Testing Deep Neural Networks}.
\newblock \bibinfo{journal}{\emph{CoRR}}  \bibinfo{volume}{abs/1803.04792}
  (\bibinfo{year}{2018}).
\newblock
\showeprint[arxiv]{1803.04792}
\urldef\tempurl%
\url{http://arxiv.org/abs/1803.04792}
\showURL{%
\tempurl}


\bibitem[\protect\citeauthoryear{Sun, Wu, Ruan, Huang, Kwiatkowska, and
  Kroening}{Sun et~al\mbox{.}}{2018b}]%
        {sun2018concolic}
\bibfield{author}{\bibinfo{person}{Youcheng Sun}, \bibinfo{person}{Min Wu},
  \bibinfo{person}{Wenjie Ruan}, \bibinfo{person}{Xiaowei Huang},
  \bibinfo{person}{Marta Kwiatkowska}, {and} \bibinfo{person}{Daniel
  Kroening}.} \bibinfo{year}{2018}\natexlab{b}.
\newblock \showarticletitle{Concolic testing for deep neural networks}. In
  \bibinfo{booktitle}{\emph{Proceedings of the 33rd ACM/IEEE International
  Conference on Automated Software Engineering}}. ACM,
  \bibinfo{pages}{109--119}.
\newblock


\bibitem[\protect\citeauthoryear{Wolpert}{Wolpert}{1996}]%
        {Wolpert_1996_NFL}
\bibfield{author}{\bibinfo{person}{David~H. Wolpert}.}
  \bibinfo{year}{1996}\natexlab{}.
\newblock \showarticletitle{The Lack of a Priori Distinctions Between Learning
  Algorithms}.
\newblock \bibinfo{journal}{\emph{Neural Comput.}} \bibinfo{volume}{8},
  \bibinfo{number}{7} (\bibinfo{date}{Oct.} \bibinfo{year}{1996}),
  \bibinfo{pages}{1341--1390}.
\newblock
\showISSN{0899-7667}
\urldef\tempurl%
\url{https://doi.org/10.1162/neco.1996.8.7.1341}
\showDOI{\tempurl}


\bibitem[\protect\citeauthoryear{Zhang, Zhang, Zhang, Liu, and Khurshid}{Zhang
  et~al\mbox{.}}{2018}]%
        {zhang2018deeproad}
\bibfield{author}{\bibinfo{person}{Mengshi Zhang}, \bibinfo{person}{Yuqun
  Zhang}, \bibinfo{person}{Lingming Zhang}, \bibinfo{person}{Cong Liu}, {and}
  \bibinfo{person}{Sarfraz Khurshid}.} \bibinfo{year}{2018}\natexlab{}.
\newblock \showarticletitle{Deeproad: Gan-based metamorphic testing and input
  validation framework for autonomous driving systems}. In
  \bibinfo{booktitle}{\emph{Proceedings of the 33rd ACM/IEEE International
  Conference on Automated Software Engineering}}. ACM,
  \bibinfo{pages}{132--142}.
\newblock


\end{thebibliography}
